\documentstyle[dina4,seceqn,ams,epsf]{elsart}

\newcommand{\Tr}{\mbox{Tr}}
\newcommand{\intP}{\mbox{Pr} \!\!  \int}
\newcommand{\be}{\begin{equation}}
\newcommand{\ee}{\end{equation}}
\newcommand{\la}{\label} 
\newcommand{\bea}{\begin{eqnarray}}
\newcommand{\eea}{\end{eqnarray}}
\newcommand{\g}{G}
\newcommand{\f}{F}
\newcommand{\gx}{\g \Xi}
\newcommand{\gt}{\tilde{\g}}
\newcommand{\qt}{\tilde{Q}}
\renewcommand{\v}{V^{(1)}}
\newcommand{\vv}{V^{(2)}}
\newcommand{\ff}{(\f|\f)^{-1}}
\newcommand{\oo}{\Omega^{(r)}}
\newcommand{\Gt}{\Gamma(t)}
\newcommand{\xt}{\tilde{\Xi}}
\newcommand{\xd}{\dot{\Xi}}
\newcommand{\gv}{\vec{\g}}
\newcommand{\ga}{\g^{\alpha}_{q, \nu}}
\newcommand{\gaq}{\g^{\alpha}_q}
\newcommand{\bt}{\tilde{\beta}}

\newcommand{\gam}{\g^{\alpha}_{-q}}
\newcommand{\dsz}{\delta S^z}

\newcommand{\shm}{\hat{S}^{-}}
\newcommand{\shp}{\hat{S}^{+}}
\newcommand{\shpm}{\hat{S}^{\pm}}
\newcommand{\gtrsim}{\tilde{>}}
\newcommand{\gtrless}{>/<}
\begin{document}
\begin{frontmatter}
\title{Selfconsistent Approximations in Mori's Theory}
\author{G. Sauermann}, 
\author{H. Turschner}, 
\author{W. Just\thanksref{DANK}}
\address{Theoretische Festk\"orperphysik,
Technische Hochschule Darmstadt,
Hochschulstra\ss e 8, D--64289 Darmstadt, Germany,\\
e--mail: wolfram@arnold.fkp.physik.th-darmstadt.de} 
\thanks[DANK]{supported by Deutsche Forschungsgemeinschaft}
\begin{abstract}
The constitutive quantities in Mori's theory,
the residual forces, are expanded
in terms of time dependent correlation functions
and products of operators at $t=0$, where it is
assumed that the time derivatives of the observables
are given by products of them. 
As a first consequence the Heisenberg dynamics 
of the observables are obtained as an expansion of the same type.
The dynamic equations for correlation functions result to be
selfconsistent nonlinear equations of the type known from 
mode--mode coupling approximations. The approach yields a neccessary
condition for the validity of the presented equations. As a third
consequence the static correlations can be calculated from
fluctuation--dissipation theorems, if the observables obey
a Lie algebra. For a simple spin model the convergence of the
expansion is studied. As a further test, dynamic and static
correlations are calculated for a Heisenberg ferromagnet at low
temperatures, where the results are compared to those of a
Holstein Primakoff treatment.
\end{abstract}
\end{frontmatter}
\newpage
\section{Introduction}
In Mori's theory \cite{1} the dynamics of a set of relevant observables in the 
Heisenberg picture are transformed to an equation of motion of the
Langevin type showing a systematic part and a residual (stochastic) force as it
is suggested by the phenomenological theory of Brownian motion. The great
success of Mori's theory is due to the fact that the rewritten equations of
motion allow for excellent approximations in calculating linear response 
functions which are determined by the systematic part of Mori's equations.

In treating the systematic part of the equation the main problem is to get
an adequate approximation for the integral kernel given by the correlation
functions of the residual forces. For its evaluation mainly two methods have
been very successful: A simple perturbation theoretical
calculation (for an introduction see \cite{2}) 
and a factorization procedure into correlation functions of
the observables (mode--mode coupling, e.g. \cite{3,4} and references
therein).

From a general point of view, the basic quantity in Mori's theory is the
residual force. The purpose of the present paper is to point out that for a
large class of systems of interacting particles or spins, one can find a
systematic approach for the residual force itself, thus being able to go
beyond the calculation of linear response functions in terms of
static correlation functions: From the residual force one can 
deduce approximations for
\begin{itemize}
\item[i)] the Heisenberg dynamics of the relevant operators
\item[ii)] the dynamic correlation functions of the observables
\item[iii)] the static correlations
\end{itemize}

We will show that for systems as interacting spins, where the time
derivatives of the relevant observables $G$ are given by a superposition of
products of the observables
\begin{equation}
\label{equ-i-1}
\left[ {\cal{H}}, G \right] = V^{(1)} G + V^{(2)} GG + V^{(3)} GGG + \cdots
\end{equation}
one can expand the residual force in terms of the dynamic correlation 
functions. This leads to an approximation scheme 
for the points (i) and (ii), and
together with a Lie algebra for the observables to point (iii). In using this
scheme one can see that the calculated Heisenberg dynamics do not 
have divergent secular terms which occur in a simple perturbation treatment.
The dynamics of the correlation functions are governed by a set of nonlinear 
equations which are close to the usual mode--mode 
coupling equations. The static
correlations in these equations can be expressed by dissipation--fluctuation 
relations 
and the Heisenberg dynamics which are given by the dynamic correlation 
functions 
and operators at time $t=0$. Thus all quantities in principle can be calculated
from the bare interactions $V^{(1)}, V^{(2)},\ldots $.

As already mentioned the essential condition for our treatment is that
the time derivatives $ \left[ {\cal{H}}, G \right]$ of the relevant observables
are given by the products of the observables (Eq.(\ref{equ-i-1})). 
Additionally we exploit the
unitarity of the time evolution in Hilbert space which written with the
Liouvillian
\begin{equation}
\label{equ-i-2}
\exp (iLt) \, (GG) = \left(  \exp (iLt) \, G \right)  \left(  \exp 
(iLt) \, G \right)\quad ,
\end{equation}
is not automatically included in a Liouville space calculation, since $G$
and $GG$ are just two different elements of the space.

It is to be expected that our expansion scheme could be generalized to the
case where in Eq.(\ref{equ-i-1}) additional 
contributions have to be taken into account which are not 
products of the chosen set.
But this extension is beyond the scope of our paper. Our
principal interest is to show that besides of linear response, Mori's
theory provides a powerful tool to find selfconsistent approximations for
the time dependence of the Heisenberg operators $G(t)$, thus leading to
the possibility of calculating higher order response functions or expectation
values for relaxation processes beyond linear response.

Our paper is organized as follows: In section 2, after having given a short
summary of Mori's theory, we present a formal expansion of the residual forces
in terms of the dynamic correlation functions, and then discuss the 
approximation
scheme for the points (i)--(iii).

In section 3 we want to illustrate the formal results of section 2 
from different
points of view. First we will study the accuracy of the approximations by
treating a simple model which can be solved exactly. Secondly, we will show
how the formalism can be applied to a physical system, and will compare
the results to other approaches.
\section{Expansion into powers of correlation functions}
\subsection{Summary of Mori's theory}
To summarize Mori's theory we take the notation of \cite{2}. The theory starts
with a scalar product in Liouville space given by
\begin{equation}
\label{equ-ii-1}
(A|B) = \beta^{-1} \int\limits^{\beta}_{0} d\lambda \langle A^\dagger 
B(i \lambda) \rangle _\beta \quad ,
\end{equation}
where $B(t)$ denotes the time evolution in the Heisenberg picture. 
Projecting the observables
$G_{\mu}(t)$ onto the space spanned by all the $G_{\nu}$ with
\begin{equation}
\label{equ-ii-2}
P \, G_{\nu} = (1-Q) G_{\nu} = G_{\nu} \quad ,
\end{equation}
and onto the orthogonal space, one obtains the Langevin operator for the
column vectors $\dot{G}(t)$ in the form
\begin{equation}
\label{equ-ii-3}
\dot{G}(t) = i \, G(t)  \Omega \, - \, \int\limits^{t}_{0} \, G(t-t') \,
\, (G|G)^{-1} (f|f(t'))dt'+ f(t) \quad ,
\end{equation}
where $f(t)$ is the column vector of the residual forces with the property
\begin{equation}
\label{equ-ii-4}
Qf(t) = f(t) \quad .
\end{equation}
The abbreviations mean
\begin{equation}
\label{equ-ii-5} 
(G(t)   \Omega)_{\mu} = \sum\limits_{\nu}  G_{\nu}(t)  \Omega_{\nu \mu}
\end{equation}
\begin{equation}
\label{equ-ii-6}
\left( (G|G)^{-1}  (f|f(t)\right)_{\nu \mu} = \sum\limits_{\lambda} 
(G|G)^{-1}_{\nu \lambda} 
(f_\lambda|f_\mu(t)) \quad ,
\end{equation}
and the frequency matrix is given by the Liouville operator $L$
\begin{equation}
\label{equ-ii-7}
\Omega_{\nu \mu} = \sum\limits_{\lambda} (G|G)^{-1}_{\nu \lambda}
(G_\lambda |L G_\mu) \quad .
\end{equation}
The main point is that the dynamics of the residual forces $f(t)$ are
not governed by the Liouvillian, but by $QLQ$.

The normalized correlation matrix $\Xi(t)$
\begin{equation}
\label{equ-ii-8}
\Xi_{\nu \mu} (t) = \sum\limits_{\lambda} (G|G)^{-1}_{\nu \lambda}  
(G_\lambda|G_\mu(t)) 
\end{equation}
obeys the matrix equation
\begin{equation}
\label{equ-ii-9}
\dot{\Xi}(t) = i  \Xi(t) \Omega - \int\limits^t_0  \Xi(t-t')(G|G)^{-1}
(f|f(t')) dt' \quad .
\end{equation}
With help of the correlation matrix $\Xi(t)$, the Langevin equations (\ref{equ-ii-3})
can formally be integrated to yield
\begin{equation}
\label{equ-ii-10}
G(t) = G  \Xi(t) + \int\limits^t_0  f(t')  
\Xi (t-t') dt' \quad .
\end{equation}
It is this form which we will use in our following treatment.
\subsection{Expansion of $f(t)$ in terms of $\Xi(t)$ and static correlations}
To find an expansion for the residual force we first establish an exact
nonlinear system of equations for $f(t)$ in terms of the correlation
functions (\ref{equ-ii-8}). This will be achieved in two steps: 
We express $f(t)$ by
the time derivatives $\dot{G}(t)$, insert Eq.(\ref{equ-i-1}) which 
leads to products
of $G(t)$. Then the decomposition (\ref{equ-ii-10}) of all $G_\nu(t)$ 
is used to get
a closed system for $f(t)$.

Take the time derivative of Eq.(\ref{equ-ii-10}), then 
using Eqs. (\ref{equ-ii-2}) and (\ref{equ-ii-4}) we find
\begin{equation}
\label{equ-ii-11}
f(t) = Q \, \dot{G}(t) - f(t)  \otimes  \dot{\Xi}(t) \quad ,
\end{equation}
where we abbreviate the convolution of two functions $a(t)$ and $b(t)$ by
\begin{equation}
\label{equ-ii-12}
\int\limits^{t}_{0}  a(t-t')  b(t') dt' = a(t)  \otimes  b(t)  \quad .
\end{equation}
Substituting the derivatives (\ref{equ-i-1}) at time $t$
\begin{equation}
\label{equ-ii-13}
[{\cal{H}}, G_\mu (t)] = \sum\limits_{\nu}  V^{(1)}_{\mu, \nu}  
G_\nu (t) + \sum\limits_{\nu, \lambda}  V^{(2)}_{\mu, \nu \lambda} 
G_\nu (t)  G_\lambda (t) + \cdots
\end{equation}
abbreviated by
\begin{equation}
\label{equ-ii-14}
[{\cal{H}}, G(t)] = V^{(1)}  G(t) + V^{(2)} 
\left\{G(t), G(t)\right\} + \cdots 
\end{equation}
into Eq.(\ref{equ-ii-11}) we arrive at
\begin{equation}
\label{equ-ii-15}
f(t)  =  iQ  V^{(2)} \left\{ G(t), 
 G(t) \right\} - i f(t)  \otimes (G|G)^{-1}
(G|V^{(2)} \left\{ G(t) , G(t) \right\} )+ \cdots \quad ,
\ee
where use has been made of the definition of $\dot{\Xi}(t)$ (\ref{equ-ii-8}), 
and of the relation
\begin{equation}
\label{equ-ii-16}
V^{(1)} \, Q(G(t) - f(t)  \otimes  \Xi(t)) = V^{(1)}  Q ( G 
\Xi(t)) = 0
\end{equation}
which follows from Eqs.(\ref{equ-ii-10}) and (\ref{equ-ii-2}).

Finally we insert the decomposition (\ref{equ-ii-10}) of $G(t)$ into 
Eq.(\ref{equ-ii-15}). Then a closed set of equations arises which implicitely 
defines $f(t)$ in terms $\Xi(t)$ and static correlations. To make it
clear we write down the result for the case, where $V^{(3)}, V^{(4)} \cdots$
vanish. One yields
\begin{eqnarray}
\label{equ-ii-17}
f(t) & = & iQ V^{(2)} \left\{G  \Xi(t) + f(t)  \otimes \Xi(t),
G \Xi(t) + f(t) \otimes  \Xi(t) \right\}  \\ 
& - & i f(t) \, \otimes  (G|G)^{-1} (G|V^{(2)}
\left\{G  \Xi(t) 
+  f(t)  \otimes  \Xi(t), G  \Xi(t) + f(t)
\otimes \Xi(t) \right\} )\nonumber \quad .
\end{eqnarray}
For simplicity we restrict to this case in the following.

The implicit system (\ref{equ-ii-17}) for $f(t)$ is our basic result and
will be the starting point for our approximations. 
To find an explicit expression
for $f(t)$ we will iteratively solve Eq.(\ref{equ-ii-17}). Let us illustrate
this procedure for a simple case.

The expansion to be chosen depends on the magnitude of the matrix elements
of $V^{(2)}$. Suppose that we can take all matrix elements
of $V^{(2)}$ to be of the same order of magnitude of a smallness parameter
$\epsilon$,
\begin{equation}
\label{equ-ii-18}
V^{(2)}   \sim  \epsilon \quad .
\end{equation}
Then we can write
\begin{equation}
\label{equ-ii-19}
f(t) = \epsilon f^{(1)}(t) + \epsilon^2 f^{(2)}(t)  +  \cdots \quad ,
\end{equation}
and comparing both sides of Eq.(\ref{equ-ii-17}) we will find the 
explicit expressions
for $f^{(n)}(t)$ in terms of correlation functions $\Xi(t)$ and 
products of operators $G$.

The lowest order contribution yields
\begin{equation}
\label{equ-ii-20}
f(t) = iQ \, V^{(2)} \left\{ G  \Xi(t),  G  \Xi(t)\right\} 
+ \cdots \quad  ,
\end{equation}
where $f(t)$ lies in the space spanned by $G$ and the products $GG$, and the
time--dependent coefficients are given by products of $\Xi(t)$.

From a physical point of view the result (\ref{equ-ii-20}) can be 
interpreted as a sum of
coupled modes, whose time dependency is given by $\Xi(t)$. 
In Eq.(\ref{equ-ii-20})
the projector $Q$ acting on the products $GG$ projects out all contributions
of the linear space spanned by the modes $G_\nu$. Therefore the effect 
of the coupling
of the $\Xi(t)$ can be very weak, although formally the bare interaction 
strength $V^{(2)}$ appears. This point will become more clear in our example
in section 3, where we treat interacting spin waves. A further comment should
be given as to the time scale for $f(t)$ appearing in Eq.(\ref{equ-ii-20}). 
In a macroscopic
interacting system the matrix $V^{(2)}$ implies a summation over a very large
number of products $\Xi_{\nu \mu}(t) \, \Xi_{\lambda \sigma}(t)$. So the phase
factors of these products can produce a correlation time for $f(t)$ which is
entirely different from the relaxation times of $\Xi(t)$.

Iterating Eq.(\ref{equ-ii-17}) one step further one 
obtains $f^{(2)}(t)$, which 
then results to be in the space spanned by $G, GG, GGG$. 
Although mathematically
possible, we have found from the example discussed in section 3.1 that instead
of calculating $f^{(2)}(t)$ it is more adequate to take a larger set of
observables $\tilde{G}$ which comprises the products $GG$, and use the
lowest order approximation (\ref{equ-ii-20}) to this set. Then one obtains
\begin{equation}
\label{equ-ii-21}
\tilde{f}(t) = i\tilde{Q} \tilde{V}^{(2)} \left\{ \tilde{G}  \tilde{\Xi}(t),
 \tilde{G}  \tilde{\Xi}(t) \right\} \quad ,
\end{equation}
where $\tilde{f}(t)$ is the column-vector of the residual forces arising
for the set $\tilde{G}$. It includes the contribution
of $f^{(2)}$ of the original set $\{G\}$. The relation between $\tilde{f}(t)$
and $f^{(2)}(t)$ is discussed in appendix A.

The simple expansion scheme starting with Eq.(\ref{equ-ii-18}), 
or Eq.(\ref{equ-ii-20}) respectively, will be illustrated in section 3.1, 
whereas 
an expansion of $f(t)$ with
\begin{equation}
\label{equ-ii-22}
V^{(2)} = V^{(2)}_0 + \epsilon \, V^{(2)}_1
\end{equation}
will be the basis for the treatment of our example in section 3.2.
\subsection{Expansion of $G(t)$ in terms of $\Xi(t)$ and 
static correlations}
Regarding the general decomposition (\ref{equ-ii-10}) of $G(t)$ into $P
G(t)$ and $Q G(t)$ it is clear, that inserting approximations for
$f(t)$ one obtains an expansion of $G(t)$ into powers of $\Xi(t)$  and
static correlations. For the case (\ref{equ-ii-18}) the lowest order 
contribution 
follows to be
\begin{equation}
\label{equ-ii-23}
G(t) = G  \Xi(t) + iQ V^{(2)} \left\{ G \Xi(t), G \Xi (t)\right\}
\otimes  \Xi(t) + \cdots \quad .
\end{equation}
The main point of this result is, that for an interacting system, $\Xi(t)$
exhibits damping, so that the Fourier transforms of $G(t)$ do not 
have divergent denominators. Furthermore the projector $Q$ again weakens the 
effect of the given interaction $V^{(2)} \left\{ G, G \right\}$, 
as the part $P(GG)$ is already
contained in the dynamics of $\Xi(t)$. The structure of the result 
(\ref{equ-ii-23}) is 
the same as it is known from an ordinary expansion of $G(t)$ with respect to
$V^{(2)}$. For $[ {\cal{H}}^{(0)}, G] = V^{(1)} \, G$, we would find
Eq.(\ref{equ-ii-23}) without $Q$, i.e. the full interaction, 
and $\Xi(t)$ being
replaced by the transposal of exp $[i \, V^{(1)} t]$.

Once an approximation of $\Xi(t)$ is known, one can use 
Eq.(\ref{equ-ii-23}) to
calculate expectation values for relaxation processes or higher order response
functions. If one wants to improve the result (\ref{equ-ii-23}), 
one can again take the
larger set of observables $\tilde{G}$ and use Eq.(\ref{equ-ii-23}) 
for the corresponding
quantities.
\subsection{Nonlinear equations for $\Xi(t)$ in terms of static
correlations}
The correlation matrix $\Xi(t)$ is determined by Eq.(\ref{equ-ii-9}), 
i.e. by the correlations 
$(f|f(t))$. Therefore use can be made of the
expansion procedure for the residual force $f(t)$ of section 2.2 
leading to closed nonlinear
equations for $\Xi(t)$. Taking the case (\ref{equ-ii-18}) and 
substituting the lowest
order contribution of $f(t)$ (\ref{equ-ii-20}) into Eq.(\ref{equ-ii-9}) 
yields the following
set of equations
\bea
\label{equ-ii-24}
\dot{\Xi}(t) &=& i  \Xi(t) \Omega \nonumber\\
&-& \int\limits^{t}_{0} \Xi(t - t')
(G|G)^{-1} (V^{(2)} \left\{ G,G \right\}
|Q  V^{(2)} \left\{ G \Xi(t'), G \Xi(t') \right\} )  dt' \quad .
\eea

It is clear that higher order equations can be obtained, if higher order
terms of $f(t)$ are inserted into $(f|f(t))$. So we have found a systematic
way to generate equations for correlation functions $\Xi(t)$.

But we need not write down these higher order equations for $\Xi(t)$, 
because -- as already has been discussed in section 2.2 -- the higher order
approximations to $\Xi(t)$ are included in Eq.(\ref{equ-ii-24}), if used for 
an extended set of observables $\tilde{G} =\{G, GG - \langle GG \rangle , 
GGG - \langle GGG \rangle , \cdots .\}$.
with new parameters $\tilde{\Omega}$ and $\tilde{V}^{(2)}$ and the 
corresponding
correlation matrix $\tilde{\Xi}(t)$.

Inspecting the result, (\ref{equ-ii-24}) shows that it has the structure 
known from
mode--mode coupling approximations. In our case we have obtained 
equations --
as mentioned in the introduction --
with bare interactions $V^{(2)}$ \cite{5,6,7}.

As a check of the validity of the approximation (\ref{equ-ii-24}) 
one can take the 
time dependence $G(t)$ resulting from (\ref{equ-ii-23}) with the 
approximate $\Xi$ and calculate the correlation
\begin{equation}
\label{equ-ii-25}
\Delta(t) = (G|G)^{-1} \, \, (G(t)|G(t))
\end{equation}
which for the exact dynamics $G(t)$ is one. In the discussion of the 
example given
in section 3.1 we have found that the deviation of $\Delta$ from $1$
is directly related to the accuracy of the approximate $\Xi$. 
\subsection{Static correlations}
The equations of motion for $\Xi(t)$ Eq.(\ref{equ-ii-24}) 
contain static correlations.
They can be viewed as given parameters and can be taken from any static theory
which is available. In this section we want to point out that in many cases
however, one can close the dynamic equations (\ref{equ-ii-24}) 
by a set of equations for
the static correlations which appear in $\Xi(t)$, so that all quantities --
at least in principle -- can be determined in a selfconsistent way. The key
point is that the approximations obtained for the observables $G(t)$ in the
Heisenberg picture are expressed in terms of $c$--number time functions and
products of operators $G$ at time $t=0$. So the well known KMS--condition for
the Heisenberg operators $G$ at imaginary times $i\beta$
\begin{equation}
\label{equ-ii-26}
\langle G \, A \rangle _\beta = \langle A  G (i\beta) \rangle _\beta
\quad ,
\end{equation}
which can be cast into the relations
\begin{eqnarray}
(A|G) &=& \int\limits^{\infty}_{-\infty} \frac{d\omega}{\beta \omega}
\langle [G(\omega), A^{\dagger} ] \rangle _\beta \label{equ-ii-27}\\
\langle A \, G \rangle _\beta &=& \int\limits^{\infty}_{-\infty}  d\omega 
(e^{\beta \omega}-1)^{-1}
\langle [G(\omega), A] \rangle _\beta \quad , \label{equ-ii-28}
\end{eqnarray}
directly connects static and dynamic correlations. Here $G(\omega)$ denotes
\begin{equation}
\label{equ-ii-29} 
G(\omega) = \frac{1}{2\pi}  \int\limits^{\infty}_{-\infty}  G(t)
e^{i\omega t} dt \quad ,
\end{equation}
and it has been assumed that the observables $G$ are chosen to be 
orthogonal to 1
and no other constants of motion are to be projected out.
Therefore we must only show that given an approximation for $G(t)$, or $f(t)$
respectively, Eq.(\ref{equ-ii-27}) and Eq.(\ref{equ-ii-28}) will 
provide us with a closed set of 
equations for the relevant static correlations appearing in $\Xi(t)$. To
this end we will suppose that the operators $G$ obey a Lie-
algebra\footnote{The matrix $C$ can originate from the fact that the 
$G$ have been chosen with the property 
$(1|G)_\beta = \langle G \rangle _\beta = 0$.}\footnote{If 
non--Hermitian operators $G$ are used, the 
operators $G^{\dagger}$ are
assumed to be elements of the space spanned by the set $G$.}
\begin{equation}
\label{equ-ii-30}
[G, G] = \sum  \cdots  G + C1
\end{equation}
leading to
\begin{eqnarray}
\left[ G, G G \right] &=& \sum \, \cdots \, G G + \sum \ldots G
\label{equ-ii-31} \\
\left[ G G , G G \right] &=& \sum \, \cdots \, G G G + \sum \ldots G G
\quad .
\label{equ-ii-32}
\end{eqnarray}
For clarity in the presentation we refrain from writing down the structure
constants explicitly, but focus on the observables.
Let us first discuss the expansion scheme $V^{(2)} \sim \epsilon$ of
section 2.2 
with the lowest order approximation for $G(t)$ found to be
\begin{equation}
\label{equ-ii-33}
G(t) = G  \Xi + i  QV^{(2)} \left\{ G  \Xi ,  G  
\Xi \right\}
\otimes \Xi \quad .
\end{equation}
The corresponding equation for $\Xi$ is given by (\ref{equ-ii-24}). 
If we apply the Kubo
identity
\begin{equation}
\label{equ-ii-34}
\beta (G|LA)_\beta = \langle [G^{\dagger}, A] \rangle _\beta
\end{equation}
to its frequency and memory matrices and carry out the commutators with
help of Eqs.(\ref{equ-ii-30}) and (\ref{equ-ii-31}), 
and $\langle G \rangle _\beta = 0$, then we see that
$\Xi(t)$ depends on the static correlations $(G| G)$, 
$\langle G G \rangle$, $(G| GG)$ only
\begin{equation}
\label{equ-ii-35}
\Xi = \Xi(t, (G|G), \langle GG \rangle , (G|GG)) \quad .
\end{equation}

To connect these static quantities we start with  two relations which
are independent of the approximations. Regarding 
$LG = V^{(1)}G + V^{(2)}\left\{ G, G \right\}$ and calculating 
$(G|LG)$ from the Kubo identity with Eq.(\ref{equ-ii-30}) we arrive at
\begin{equation}
\label{equ-ii-36}
C = \beta (G|V^{(1)}G) + \beta(G|V^{(2)}\left\{ G, G\right\} ) \quad .
\end{equation}
On the other hand we calculate $\langle G^{\dagger} G \rangle $ applying 
Eq.(\ref{equ-ii-28}) to $A = G^{\dagger}$. Inserting
(\ref{equ-ii-33}) and the equation of motion for $\Xi(\omega)$ we arrive at
\begin{equation}
\label{equ-ii-37}
\langle G^{\dagger} 
G \rangle  = (G|G)  \int\limits^{\infty}_{- \infty}  d\omega 
\frac{\beta \omega}{e^{\beta \omega} -1} \Xi (\omega) \quad .
\end{equation}

Two further relations are obtained, if we explicitly use the approximation
(\ref{equ-ii-33}) for $G(t)$ or $G(\omega)$ respectively. 
We regard Eq.(\ref{equ-ii-27}) 
for $A = (GG)^{\dagger}$ and insert the Fouriertransform of 
(\ref{equ-ii-33})\footnote{The correlations $(G|G)$ cannot be calculated from
Eq.(\ref{equ-ii-27}) with $A=G$, as in this case Eq.(\ref{equ-ii-27})
identically holds, if the equation of motion for $\Xi(\omega)$ is
inserted.}.
Then respecting the commutators (\ref{equ-ii-31}) and (\ref{equ-ii-32})
we find equations of the following structure 
\begin{equation}
\label{equ-ii-38}
(G|GG) = \sum \cdots \langle GG \rangle  + 
\sum  \cdots 
\langle GGG \rangle  + \sum \cdots  \langle GG \rangle  (G|G)^{-1} (G|GG)
\quad ,
\end{equation}
where we have omitted the coefficients 
which are given by the structure constants of the Lie algebra and matrix
elements of
\begin{equation}
\label{equ-ii-39}
\int \, \frac{d\omega}{\beta \omega} \Xi(\omega) \quad \mbox{and} \quad 
\int \, \frac{d\omega}{\beta \omega} (\Xi  \Xi  \otimes  \Xi)(\omega) 
\quad .
\end{equation}
To close this system of equations with 
$\langle GGG \rangle $ we regard Eq.(\ref{equ-ii-28}) for
$A = GG$ yielding
\begin{equation}
\label{equ-ii-40}
\langle GGG \rangle  =  \sum \cdots 
\langle GG \rangle  + \sum \cdots 
\langle GGG \rangle  + \sum \cdots 
\langle GG \rangle (G|G)^{-1}(G|GG)
\end{equation}
with coefficients analogous to the coefficients 
appearing in Eq.(\ref{equ-ii-38}). The
only difference is, that the denominators 
$\beta \omega$ in Eq.(\ref{equ-ii-39}) are
replaced by $(e^{\beta \omega}-1)$.

These four equations (\ref{equ-ii-36}), (\ref{equ-ii-37}), 
(\ref{equ-ii-38}), (\ref{equ-ii-40}) 
for $(G|G), \langle GG \rangle , (G|GG), \langle GGG \rangle $
now in principle allow for a determination of the static
correlations in terms of functionals of the dynamic correlations $\Xi(\omega)$,
where it is clear, that the solvability must be discussed in each
application.
Although these equations are extremely complicated, one can 
think of an iterative solution starting with 
suitable static values in the dynamics of $\Xi$.  Another 
simplification occurs, 
if the temperature is high enough so that $\tanh \beta \omega$ can be 
replaced by $\beta \omega$.
Then it holds
\begin{eqnarray}
\label{equ-ii-41}
(G|G) &=& \frac{1}{2} \langle [G^{\dagger}, G]_+ \rangle  \\
(G|GG) &=& \frac{1}{2} \langle [G^{\dagger}, GG]_+ \rangle \quad ,  
\end{eqnarray}
and the total system reduces to a set for the correlations 
$\langle GG \rangle $ and
$\langle GGG \rangle $.

We emphasize that the basis for our discussions on the statics was the 
result (\ref{equ-ii-33}) for the Heisenberg
dynamics. If we now consider higher order contributions to $G(t)$, we have
found two possibilities, as was pointed out in section 2.2 and 2.3. But as 
to the static correlations these two treatments are not equivalent.
If we improve Eq.(\ref{equ-ii-33}) by a residual force $f^{(2)}(t)$, so 
that $G(t)$ is an element
of the space spanned by $G, GG, GGG$, then it is not difficult to see, that
the static equations resulting from (\ref{equ-ii-27}) and (\ref{equ-ii-28}) 
are no longer closed.
However, if we use the alternative and extend the set of observables to yield
a set $\tilde{G}$ which includes the products $GG$, then the set of equations
for the static correlations appearing in ${\tilde{\Xi}}(t)$ again can be 
closed
(appendix A). This also indicates that the second way is more adequate.

So far we have discussed the equations for the static correlations which
result from the expansion scheme for the residual forces with 
$V^{(2)} \sim \epsilon$.
In general the static relations to be obtained will depend on the
iteration procedure for $f(t)$ which will be chosen. In our example of section
3.2, where we treat the lowest  order of a case $V^{(2)} = V^{(2)}_0 +
\epsilon V^{(2)}_1$, we will see that a closed set of equations can be 
derived again.

The considered approaches leading to closed sets for the static correlations
have the common feature that they are valid to the same degree of accuracy
as the dynamics in the Heisenberg picture is correct. In this sense dynamics
and statics are treated at the same level of approximation.
\section{Illustrations and discussion}
In this section we will illustrate the general formalism of section 2 applying
it to two examples. First we will study general features of the approximation
scheme. To this end we choose a model as simple as possible: We take a spin 
system
with long range interactions in the high temperature limit, so that we need 
not 
handle with the problem of the static correlations, and can obtain both, the
exact solution to the dynamics and the approximations in analytic form. As
a second example we will treat a Heisenberg ferromagnet at low temperatures. We
will show that one can find an expansion of the residual forces
in terms of spin operators which leads to meaningful results for the
Heisenberg dynamics and the correlation functions, so that we can make
contact to other approaches.
\subsection{Exactly solvable model in the high temperature limit}
Our model to be considered is a Heisenberg spin system with long range
interactions in the high temperature limit, where we can carry through the 
general expansions of sections 2.2 and 2.3 to high order explicitly. We will
find that the lowest order approximation to the residual forces leads to a
coupling of two modes only, where one of them is a trivial constant of the
motion. So the system might be a very special one, but we think that it
correctly gives insight into the problems of time scales and convergence.

The Hamiltonian of our system reads
\be
\label{equ-iii-1}
{\cal H} = - \frac{J}{\sqrt{N}} \sum_{ij} \vec{s}_i \cdot \vec{s}_j =
 - \frac{J}{\sqrt{N}} \vec{S}_0 \, \cdot \, \vec{S}_0 ,
\ee
where
\bea
\vec{S}_q &=& \sum\limits^{N}_{i=1} e^{i \vec{q}  \vec{R}_i} \vec{s}_i
\label{equ-iii-2} \\
\vec{S}_0 &=& \vec{S}_{q=0} 
\label{equ-iii-3} \quad,
\eea
and the $N$ spins  $(s=\frac{1}{2})$ are located at lattice sites
$\vec{R}_i$. In the high temperature limit the scalar product reduces to
\begin{equation}
\label{equ-iii-4}
(A|B) = (\Tr  1)^{-1} \Tr (A^{\dagger} \, B) ,
\end{equation}
so that static spin correlations can be evaluated.

In choosing a set of observables $\{ G \}$ 
we take $\vec{S}_q$, but substract
the components parallel to the constant of the motion 
$\vec{S}_0$, so that the dynamic
correlation functions $\Xi (t)$ will decay to zero (c.f. appendix B). 
Therefore we take the set of observables to be
\begin{eqnarray}
\label{equ-iii-5}
\vec{G}_0 & = & \vec{S}_0 \nonumber \\
\vec{G}_q & = & \vec{S}_q - \frac{1}{2} (\vec{S}_0 \cdot \vec{S}_0)^{-1}
\left( (\vec{S}_0 \, \cdot \,  \vec{S}_q) \vec{S}_0 + 
\vec{S}_0 (\vec{S}_0 \, \cdot \, \vec{S}_q)\right)
\quad q \neq 0
\end{eqnarray}
with $\vec{S}_0 \cdot \vec{S}_q = \sum_{\alpha} S^{\alpha}_0  S^{\alpha}_q$.
Then it holds
\begin{equation}
\label{equ-iii-6}
L \vec{G}_q = i \frac{J}{\sqrt{N}} \left( \vec{G}_q \times G_0 - G_0 
\times \vec{G}_q\right) \quad q \neq 0 \quad ,
\end{equation}
which means, that the conditions of section 2.2
\begin{equation}
\label{equ-iii-7}
L G = V^{(1)} G + V^{(2)} \{ G, G\}
\end{equation}
are fulfilled, where $V^{(1)}$ vanishes and $V^{(2)}$ just couples 
$\vec{G}_q$ to $\vec{G}_0.$ From the symmetry of the Hamiltonian it directly
follows that the correlation matrix $\Xi(t)$ is diagonal with respect to
$\vec{q}, \vec{q}\,'$ and cartesian components $\alpha, \alpha'= x, y, z$, so
that the decompositions (\ref{equ-ii-10}) specialize to
\bea
q = 0 \, &:& \,  \vec{G}_0(t) = \vec{G}_0 \, \cdot \, 1  + 0
\label{equ-iii-8} \\
q \neq 0 \, & : &\, \vec{G}_q(t) = 
\vec{G}_q \Xi(t) + \vec{f}_q(t) \otimes \Xi(t) 
\quad ,
\la{equ-iii-9}
\eea
where
\be
\la{equ-iii-10}
\Xi(t) = (G^\alpha_q | G^\alpha_q)^{-1} 
\left(G^\alpha_q | G^\alpha_q(t)\right)
\ee
for $q\neq 0$ does not depend on $q$ and $\alpha$.
So we are ready to study the expansion of the residual forces 
$\vec{f}_q(t)$ with an
expansion parameter $\epsilon \sim J$.
\subsubsection{The first order approximation}
We now take our model to carry through the expansions of section 2 for the case
$V^{(1)} =0$, $V^{(2)} \sim \epsilon \sim J$ iterating the 
Eq.(\ref{equ-ii-17}) for
$\vec{f}_q(t)$.
Although $J$ will be a formal expansion parameter the accuracy of
the approximations will not be determined by the smallness of $J$ but
rather by a condition on time $Jt \ll \delta$, as for $V^{(1)} =0$ we just
have one time scale given by $J^{-1}$. Nevertheless in a high
order approximation the bound $\delta$ may be so large that
the ''short time expansion'' covers the
whole range of physical interest. We will study this problem of time range in
detail, calculating the correlation function $\Xi(t)$ to general order.

The lowest order approximation for the residual force $\vec{f}_q(t)$ follows
from (\ref{equ-ii-20}), (\ref{equ-iii-8}) and (\ref{equ-iii-9}) 
to yield\footnote{As $(G_q|LG_q) = 0$ the projection
operator $Q$ can be ommitted.}
\be
\la{equ-iii-11}
\vec{f}_q(t) \, = \, \frac{J}{\sqrt{N}}  (\vec{G}_0 \times 
\vec{G}_q - \vec{G}_q 
\times \vec{G}_0) \Xi = iL \vec{G}_q \Xi \quad ,
\ee
where just two modes $\Xi_q = \Xi$ \, and $\Xi_{q=0} = 1$ couple, so that the
force $\vec{f}_q(t)$ for all times is proportional to one fixed element in the
space of the $GG$.

The dynamics of the Heisenberg operators $\vec{G}_q(t)$ corresponding to the
approximation (\ref{equ-iii-11}) are obtained from (\ref{equ-ii-23}) to give
\bea
\la{equ-iii-12}
\vec{G}_q(t) &=& \vec{G}_q \Xi(t) +  \frac{J}{\sqrt{N}} (\vec{G}_0 
\times \vec{G}_q - \vec{G}_q 
\times \vec{G}_0)\Xi(t) \otimes \Xi(t) \nonumber\\
&=& \vec{G}_q \Xi \, + \, iL\vec{G}_q \Xi \otimes \Xi \quad .
\eea
In this expansion the coefficient $\Xi(t)$ of $\vec{G}_q$ is exact, 
whereas the
coefficient $\Xi  \otimes  \Xi$ of $iL\vec{G}_q$ will be modified 
by higher order
terms, if these are projected onto $L\vec{G}_q$. 
The equation of motion for the
correlation function $\Xi(t)$ is given by Eq.(\ref{equ-ii-24}) 
and simplifies to
\be
\la{equ-iii-13}
\dot{\Xi} = - m_2  \Xi  \otimes  \Xi \quad ,
\ee
where
\begin{equation}
\la{equ-iii-14}
m_2 = \frac{(G^{\alpha}_q|L^2  
G^{\alpha}_q) }{(G^{\alpha}_q|G^{\alpha}_q)}
\qquad (q \neq 0)
\end{equation}
is independent of $q$ and $\alpha$.
The solution of (\ref{equ-iii-13}) for $\Xi(t)$ can be expressed by the 
Bessel function
$J_1$. Taking the Laplace transform of (\ref{equ-iii-13}) one finds
\be
\la{equ-iii-15}
\Xi^{(1)}(s) = 2 \frac{-s + \sqrt{s^2 + 4 m_2}}{4 m_2} \quad ,
\ee
or
\be
\la{equ-iii-16}
\Xi^{(1)}(t) = 2  \frac{J_1(\tau)}{\tau}  , \quad  \tau = 2 \sqrt{m_2} t
\ee
respectively, where the index denotes the order of the approximation.
This result for $\Xi^{(1)}(t)$
is compared to the exact solution in Fig.1.
\begin{figure}[t]
\epsffile{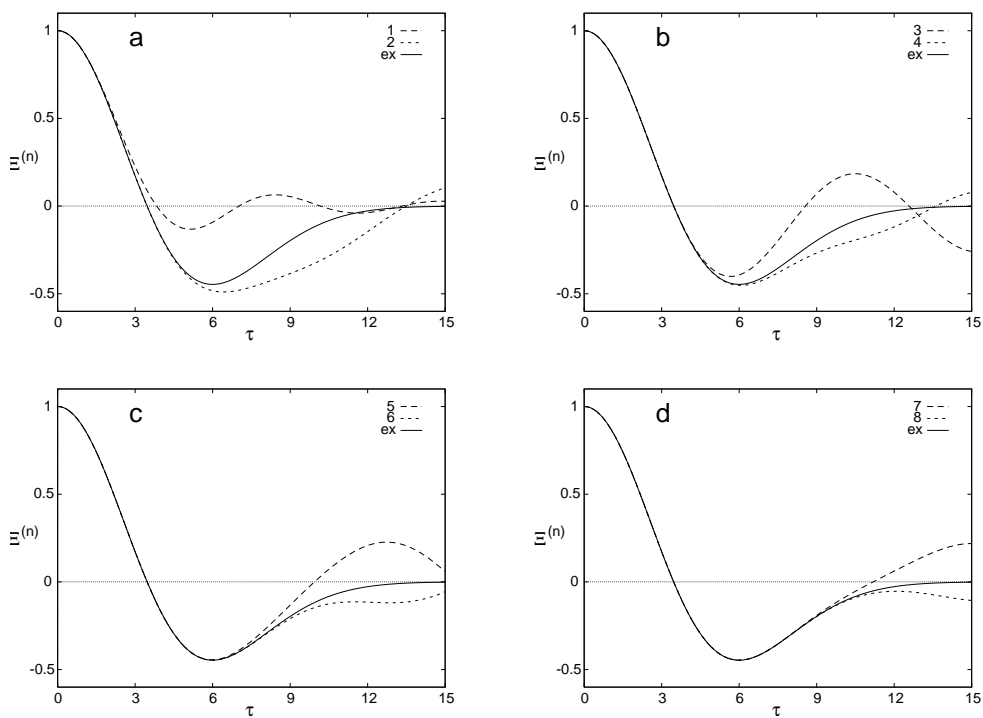}
\caption{Approximations $\Xi^{(n)}(\tau)$ compared to the exact 
correlation function $\Xi(\tau)$; a: $n=1,2$, exact;
b: $n=3,4$, exact; c: $n=5,6$, exact; d: $n=7,8$, exact.}
\end{figure}
One sees that $\Xi^{(1)}(t)$
is close to the exact correlation function, 
as long as $Jt$ is smaller than the zero of
$\Xi(t)$, i.e. as long as it holds
\be
\label{equ-iii-17}
Jt  \leq  1 \quad .
\ee
At the end of section 3.1 we will show that the range of validity
(\ref{equ-iii-17}) for $\Xi^{(1)}(t)$ can be estimated from
$\Xi^{(1)}(t)$ and (\ref{equ-iii-12}) alone, without knowledge 
of the exact solution.
\subsubsection{Higher order approximations}
As there are more possibilities to derive higher approximations we will first
study the second order approximation in some detail and then give the
general results.

Iterating Eq. (\ref{equ-ii-17}) for $\vec{f}_q(t)$ up to order $J^2$ we
find for the residual force
\begin{eqnarray}
\label{equ-iii-18}
\vec{f}_q(t) & = & \frac{J}{\sqrt{N}} \left( \vec{G}_0 \times \vec{G}_q 
- \vec{G}_q \times \vec{G}_0 \right) \Xi \nonumber\\
& + & \frac{J^2}{N} \, Q \left\{ \vec{G}_0 \times \left( \vec{G}_0 \times 
\vec{G}_q 
-  \vec{G}_q \times \vec{G}_0 \right)
-   \left( \vec{G}_0 \times \vec{G}_q - \vec{G}_q \times \vec{G}_0 
\right) \times \vec{G}_0 \right\} \Xi \otimes \Xi \nonumber \\
&=&  iL\vec{G}_q  \Xi  +  Q(iL)^2 \vec{G}_q  \Xi \otimes  \Xi
\quad ,
\end{eqnarray}
and for the Heisenberg dynamics of $\vec{G}_q(t)$
\be
\la{equ-iii-19}
\vec{G}_q(t) = \vec{G}_q  \Xi + iL\vec{G}_q  \Xi  \otimes  \Xi
+ Q(iL)^2 \vec{G}_q  \Xi  \otimes  \Xi  \otimes  \Xi \quad .
\ee
The dynamics of $\Xi(t)$, however, keeps to be $\Xi^{(1)}(t)$, as the new
added term in $\vec{f}_q(t)$ is orthogonal to $\vec{f}_q(0) = iL\vec{G}_q$,
\be
\la{equ-iii-20}
(f^\alpha_q | Q(iL)^2 \, G^\alpha_q) = 0 \quad ,
\ee
so that the memory function does not change.
To improve $\Xi^{(1)}(t)$, at least third order terms in $f^\alpha_q(t)$
have to be considered which have a non--vanishing projection 
onto $iLG^\alpha_q$.

An alternative approach to the second order result described above can be
given, if we keep the basis we have obtained for 
$\vec{f}_q(t)$ and $\vec{G}_q(t)$ in (\ref{equ-iii-19}), but use a 
formulation such that
the factor of $iL\vec{G}_q$ will not be changed by higher order terms. 
This means
that higher order terms must be constructed to be orthogonal to 
the products
$iL\vec{G}_q = J/\sqrt{N} 
( \vec{G}_0 \times \vec{G}_q - \vec{G}_q 
\times \vec{G}_0)$ (c.f. appendix A). 

For this alternative approach to the second order approximation we extend
the set of observables ${G}$ to  the larger set $\{ \tilde{G} \}$ , 
which comprises
the occuring basis vectors of products $GG$, i.e. $iL\vec{G}_q$:
\be
\la{equ-iii-21}
\{ \tilde{G} \} = \{ \vec{G}_0, \vec{G}_{q,1} = \vec{G}_q, 
\vec{G}_{q,2} = iL\vec{G}_q \} .
\ee
The corresponding projection operator onto this space is denoted by 
$\tilde{P} =
1 - \tilde{Q}$. Then the general formalism of section 2 applies to the set
$\{\tilde{G}\}$, as it holds
\begin{eqnarray}
\la{equ-iii-22}
iL\vec{G}_{q,1} & = & \vec{G}_{q,2}  \nonumber \\
iL\vec{G}_{q,2} & = & \frac{J}{\sqrt{N}} \left( \vec{G}_{0} 
\times  \vec{G}_{q,2}
- \vec{G}_{q,2} \times \vec{G}_{0} \right) \quad .
\end{eqnarray}
Thus the condition (\ref{equ-i-1}) with $\tilde{V}^{(1)} \neq 0$ and 
$\tilde{V}^{(2)} \sim J$ again are fulfilled, and the lowest order
approximation (\ref{equ-ii-20}) for the residual forces
$\left\{\tilde{f}\right\} = \left\{ \vec{f}_{q,1} , \vec{f}_{q,2} \right\}$
reads
\bea
\vec{f}_{q,1} (t) &=& 0 \la{equ-iii-23} \\
\vec{f}_{q,2}(t) &=& \frac{J}{\sqrt{N}}  \tilde{Q}  \left(\vec{G}_{0} 
\times  \vec{G}_{q,2}
- \vec{G}_{q,2} \times \vec{G}_{0} \right) \Xi_{22} \la{equ-iii-24} \quad ,
\eea
where the matrix of the correlation functions $\Xi_{\nu\mu}(t)$ is
defined by
\be
\la{equ-iii-25}
\Xi_{\nu\mu} = \frac{(G^\alpha_{q, \nu}|G^\alpha_{q, \mu}(t))}
{(G^\alpha_{q, \nu}|G^\alpha_{q, \nu})} \quad \nu, \mu = 1, 2 \quad .
\ee
Hence the Heisenberg dynamics for $\vec{G}_q(t) = \vec{G}_{q,1}(t)$
follow to be
\begin{eqnarray}
\la{equ-iii-26}
\vec{G}_q(t) = \vec{G}_{q,1}(t) & = & \vec{G}_{q,1} \Xi_{11} + \vec{G}_{q,2}
\Xi_{21} + \vec{f}_{q,2} \otimes \Xi_{21} \nonumber \\
& = & \vec{G}_q \Xi_{11} + iL\vec{G}_q \Xi_{21} + Q(iL)^2 \vec{G}_q
(\Xi_{22} \otimes \Xi_{21}) \quad .
\end{eqnarray}
One sees that the result (\ref{equ-iii-26}) for $\vec{G}_q(t)$ is spanned
by the same basis vectors as it was in Eq.(\ref{equ-iii-19}). Just the
time--dependent coefficients have changed.\footnote{If one extracts the 
residual force $\vec{f}_q(t)$ with respect to $\{G\}$ from the result
(\ref{equ-iii-26}), one finds that $\vec{f}_q(t)$ also has the same basis 
vectors as the approximation (\ref{equ-iii-18}). It is not difficult
to see, that the memory functions calculated from this $\vec{f}_q(t)$ just
lead to the exact relation $\dot{\Xi} = 
- \frac{(f^\alpha_q|f^\alpha_q(t) )}{(G^\alpha_q|G^\alpha _q)}
\otimes \Xi.$}

In order to determine $\Xi(t)$, we now write down the equations of motion for
the matrix $\Xi_{\nu\mu}(t)$ (c.f. Eq.(\ref{equ-iii-22})) 
which follow from the frequency matrix (\ref{equ-ii-7})
and the memory matrix given by $(f^\alpha_{q,2}|f^\alpha_{q,2}(t) )$.
We obtain
\begin{eqnarray}
\la{equ-iii-27}
\dot{\Xi}_{11} & = & \Xi_{12} \nonumber \\
\dot{\Xi}_{12} & = & - c_2 \Xi_{11} - c_3 \Xi_{12} \otimes \Xi_{22} \\
\dot{\Xi}_{21} & = & \Xi_{22} \\
\dot{\Xi}_{22} & = & - c_2 \Xi_{21} - c_3 (\Xi_{22} \otimes \Xi_{22}) ,
\end{eqnarray}
where $c_2 = m_2$ and $c_3 = (m_4-m_2^2)/m_2$, and $m_4$ denotes the
fourth moment.

This system can be solved by Laplace transform to yield for 
$\Xi^{(2)} = \Xi_{11}$
\be
\la{equ-iii-28}
\Xi^{(2)}(s) = \frac{ \displaystyle -\frac{c_3}{c_2}s 
- 1/2 \left(s + \frac{c_2}{s}\right) + \frac{1}{2}
\sqrt{\left(s+\frac{c_2}{s}\right)^2 + 4 c_3}}{\displaystyle
-\frac{c_3}{c_2} s^2   } \quad .
\ee
The corresponding result in time $\Xi^{(2)}(t)$ is
shown in Fig.1. As compared to $\Xi^{(1)}(t)$ the range of validity of
$\Xi^{(2)}(t)$ has obviously increased.

The alternative approach for the second order approximation can be extended
to general order. Instead of iterating the Eq.(\ref{equ-ii-17})
for $\vec{f}_q(t)$ and calculating $\vec{f}_q(t)$ up to order $J^n$, we
choose a set $\{\tilde{G}\}$ given by $\vec{G}_q, iL\vec{G}_q, \cdots 
(iL)^n \vec{G}_q$, and take the lowest order approximation for the residual
forces. The details are given in appendix B. Here we just list the results.
To write them down, it is expedient to use an orthogonal basis 
$\vec{G}_{q,\nu}$ in
the space $\left\{ \tilde{G}\right\}$ introduced by
\begin{eqnarray}
\la{equ-iii-29}
\vec{G}_{q,1} & = & \vec{G}_q \nonumber \\
\vec{G}_{q,2} & = & i L \vec{G}_{q,1} \nonumber \\
\vec{G}_{q,\nu} & = & iL \vec{G}_{q,\nu-1} + \frac{(G^\alpha_{q,\nu-1} |
G^{\alpha}_{q,\nu-1} )}{(G^{\alpha}_{q,\nu-2} | 
G^{\alpha}_{q,\nu-2})}
\vec{G}_{q,\nu-2} \qquad \nu = 3, \cdots , n \quad .
\end{eqnarray}
Then the residual forces read
\bea
\la{equ-iii-30}
\vec{f}_{q,\nu}(t) &=& 0  \qquad \nu=1, \cdots n-1 \nonumber\\
\vec{f}_{q,n}(t) &=& 
\vec{G}_{q,n+1} \Xi_{nn} \quad ,
\eea
where $\vec{G}_{q,n+1}$ is defined by (\ref{equ-iii-29}) with $\nu=n+1$.
The Heisenberg dynamics result to be
\be
\la{equ-iii-31}
\vec{G}_q(t) = \vec{G}_{q,1}(t) = \sum^{n}_{\nu=1} \, \vec{G}_{q,\nu} \, 
\Xi_{\nu1} 
+ \vec{G}_{q,n+1} \Xi_{nn} \otimes \Xi_{n1} \quad .
\ee
The equations for the correlation matrix
\be
\la{equ-iii-32}
\Xi_{\nu\mu} = \frac{\left(G^\alpha_{q,\nu}|G^\alpha_{q,\mu}(t)\right)}
{\left(G^\alpha_{q,\nu}|G^\alpha_{q,\nu}\right)} 
\ee
are found from the frequency and memory matrix. The explicit solution for the
Laplace transforms of $\Xi_{\nu1}$ and $\Xi_{nn}$ are
\bea
\Xi_{\nu1}(s) &=& \frac{\Xi(s) B_{\nu-1}(s) - A_{\nu-1}(s)}
{c_1 \cdots c_\nu} (-1)^{\nu-1}\qquad \nu = 2, \cdots, n \quad ,
\la{equ-iii-33}\\
\Xi_{nn}(s) &=& B_{n-1}(s) \Xi_{n1}(s) \quad , 
\la{equ-iii-34}
\eea
where $\Xi^{(n)} = \Xi_{11}$ is given by
\be
\la{equ-iii-35}
\Xi^{(n)}(s) = \frac{A_{n-1}}{B_{n-1}} + 
\frac {\displaystyle
-\frac{B_n}{B_{n-1}} + \sqrt{ \left( \frac{B_n}{B_{n-1}} 
\right)^2 + 4 c_{n+1}}}
{\displaystyle
2(-1)^{n-1} \frac{c_{n+1}}{c_1 \cdots c_n}  B^2_{n-1}} \quad ,
\ee
and the $A_\nu(s), B_\nu(s)$ are polynominals in $s$ defined by
\begin{equation}
\la{equ-iii-36}
\begin{array}[b]{lll}
A_\nu =  s \, A_{\nu-1} + C_\nu \, A_{\nu-2}& \quad A_0  = 0& 
\quad A_1  =  1 \\
B_\nu  =  s \, B_{\nu-1} + C_\nu \, B_{\nu-2}& \quad B_0  = 1& \quad B_1  
=  s
\end{array}
\qquad \nu = 2, \cdots , n 
\end{equation}
with
\bea
\la{equ-iii-37}
c_1 &=& 1 \nonumber\\
c_\nu &=& \frac{\left(G^\alpha_{q,\nu}|G^\alpha_{q,\nu}\right)}
{\left(G^\alpha_{q,\nu-1}|G^\alpha_{q,\nu-1}\right)}\qquad
\nu = 2, \cdots n, \quad  \alpha = x, \mbox{ or } y, z \quad .
\eea
The functions $\Xi^{(n)}(s)$ for complex $s$ are
holomorphic for $\mbox{Re} s>0$ and have the correct property
\be
\la{equ-iii-38}
\mbox{Re} \Xi^{(n)} (s) \geq 0,\quad  \mbox{Re} s \geq 0 \quad .
\ee
The results for $\Xi^{(n)}(t)$ and the exact solution $\Xi(t)$ are plotted
in Fig.1. One sees that the time region, where
$\Xi^{(n)}(t)$ is a good approximation to $\Xi(t)$ increases 
with $n$. The spectral densities $1/2 \int^{\infty}_{-\infty} 
dt \Xi(t) e^{-i\omega t} = \mbox{Re} \Xi^{(n)}
(s = i\omega)$ are shown in 
Fig.2\footnote{If one wants to get approximations
which are valid for all times $0 \leq Jt \leq \infty$ one can 
average the spectral
densities $\mbox{Re} \Xi^{(n)}(i \omega)$ 
with a width $\delta$ which smoothens the 
intervals $\mbox{Re} \Xi^{(n)}(i \omega) = 0$. 
An example for $n=12$, $1/\pi 
\int d\omega' \mbox{Re}\Xi^{(n)} (i\omega')  
\delta/[(\omega-\omega')^2 + \delta^2] 
= \mbox{Re} \Xi^{(n)} (\delta + i\omega)$ 
is given in Fig.2.}. 
\begin{figure}[t]
\epsffile{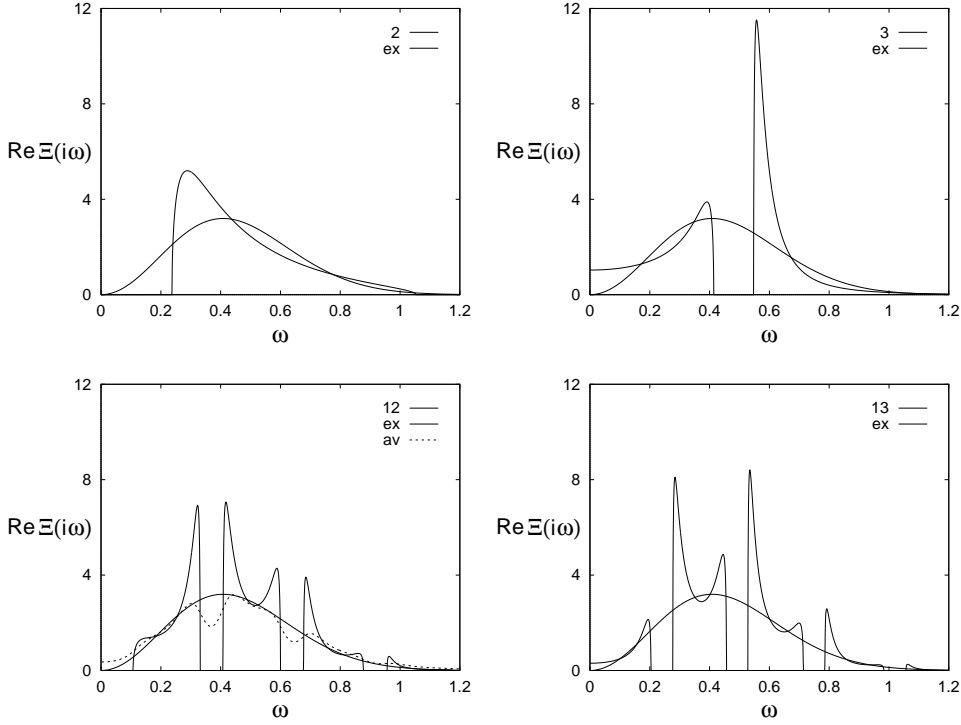}
\caption{Spectral densities of the approximations
$\mbox{Re}\Xi^{(n)}(i\omega)$, compared to the exact density for
$n=2,3,12,13$.
An averaged spectrum is shown for $n=12$. Frequency and
spectrum in units of 
$(4 m_2)^{1/2}$ and
$(4 m_2)^{-1/2}$ respectively.}
\end{figure}
The intervals where $\mbox{Re} \Xi^{(n)} (i \omega)$
identically vanishes are due to the square root occuring in $\Xi^{(n)}(s)$ 
and 
are a consequence of the coupling of just two modes. Therefore in the case
of many modes one can expect that the values 
$\mbox{Re} \Xi^{(n)} (i\omega) \neq 0$
are no longer restricted to finite intervals which cause the asymptotic
oscillations for long times in $\Xi^{(n)} (t)$.

At the end of this section we will discuss the validity of the approximations
from a different point of view. Our approach does not only lead 
to equations for
the correlation functions. The basis was the expansion of the residual force
and the resulting Heisenberg dynamics $\vec{G}_q(t)$. From this we will be able
to write down a necessary condition for the
accuracy of the $n$--th order approximation for $\Xi(t)$ and 
$\vec{G}_q(t)$.
Let us introduce the correlation at equal times
\be
\la{equ-iii-39}
\Delta(t) = \frac{(G^{\alpha}_q(t)|G^{\alpha}_q(t))}
{(G^{\alpha}_q|G^{\alpha}_q)} 
\ee
which for the exact $\vec{G}_q(t)$ must be one. 
Inserting $\vec{G}_q(t)$ from 
(\ref{equ-iii-31})
we find
\be
\la{equ-iii-40}
\Delta^{(n)}(t) = \sum^{n}_{\nu=1}  \Xi_{1\nu}(t)  \Xi_{\nu1}(t) +
c_1 c_2 \cdots \cdot c_{n+1} (\Xi_{nn}  \otimes  \Xi_{n1})^2
\ee
where the orthogonality of the $\vec{G}_{q,\nu}$ and the definitions 
(\ref{equ-iii-32})
and (\ref{equ-iii-37}) have been used. The expresion (\ref{equ-iii-40}) 
for $\Delta^{(n)}$
can be simplified further with help of the equations of motion for
$\Xi_{\nu\mu}(t)$, or the explicit solutions (\ref{equ-iii-33}) and
(\ref{equ-iii-34}), e.g.
\be
\la{equ-iii-41}
\Delta^{(1)}(t) = \left[ \Xi^{(1)}(t) \right]^2 + \frac{1}{c_2} 
\left[ \dot{\Xi}^{(1)}(t) \right]^2 \quad .
\ee
If the $n$--th order approximation is appropriate we must have
\be
\la{equ-iii-42}
\Delta^{(n)}(t) \sim 1
\ee
which gives a restriction to the region of time, as for $t=0$ it holds
$\Delta^{(n)}(t=0) = 1$.
The deviation of $\Delta^{(n)}(t)$ from 1 in Eq.(\ref{equ-iii-40})
comes from the finite basis in Liouville space entering into 
$\vec{G}_q(t)$ and
the approximation for the correlation functions, which is related to this
subspace. Thus $\Delta^{(n)}(t) \sim 1$ requires that the restricted 
Liouville space in $\vec{G}_q(t)$ as well as the approximate $\Xi^{(n)}(t)$
are sufficient. In Fig.3  
\begin{figure}[t]
\epsffile{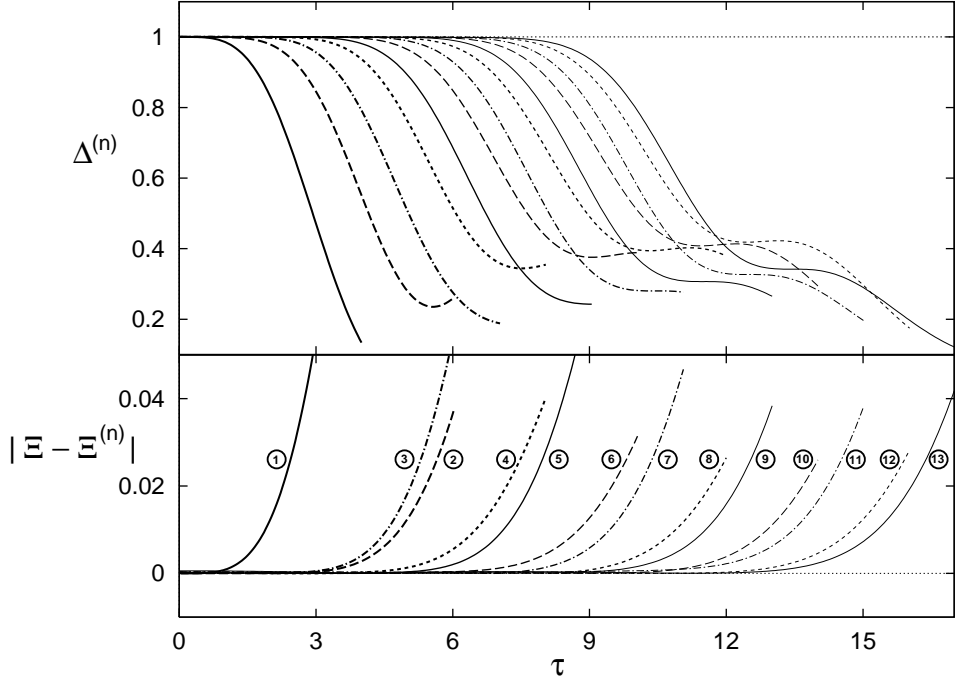}
\caption{Accuracy of the approximations; norm 
$\Delta^{(n)}(\tau)$ compared
to the deviations of $\Xi^{(n)}(\tau)$ from the exact correlation
function $\Xi(\tau)$ for several orders of the approximation
$n=1,2,\ldots,13$.}
\end{figure}
we have plotted $\Delta^{(n)}(t)$. Comparing
the deviations $\Xi^{(n)}(t)$ from $\Xi(t)$ to $\Delta^{(n)}(t)$ one
observes that $\Xi^{(n)}(t)$ is a good approximation to $\Xi(t)$
\be
\la{equ-iii-44}
\Xi^{(n)}(t)  \sim  \Xi(t) \quad ,
\ee 
as long as it holds 
\be
\la{equ-iii-45}
\Delta^{(n)}(t) \gtrsim \frac{1}{2} \quad .
\ee
From this we conclude that one can take the condition (\ref{equ-iii-42})
as a measure for the accuracy of the approximations without knowing the
exact solution $\Xi(t)$.
\subsection{Heisenberg ferromagnet at low temperatures}
Introducing the Heisenberg ferromagnet we want to show how our general
method can be applied to a system where temperature dependent static 
correlations
occur. We will not present more refined results, but rather want to study the
selfconsistent method by comparing the results to those of other approaches.
To this end we will simplify our equations for the dynamic and static 
correlation
functions, so that we can recover standard results for temperature dependent
 spin wave frequencies and dampings. In applying our formalism we keep the
spin operators and do not transform to Bose operators. Thus the Heisenberg
dynamics of the spin operators are expressed in a basis of spin operators at
$t=0$ which may have some advantage for further use.

The Hamiltonian of our system is given by
\begin{eqnarray}
\la{equ-iii-46}
{\cal H} & = & - H \sum_{i} s^z_i - \sum_{i \neq j} J_{ij} \vec{s}_i \cdot
\vec{s}_j
\nonumber\\
& = & -H S^z_0 - 1/N \sum_q J_q  \, \vec{S}_q \cdot \vec{S}_{-q}
,\quad  H \rightarrow + 0, \quad J_{ij} > 0 
\end{eqnarray}
where $N$ spins $S$ are located at sites $\vec{R}_i$ in a
lattice and $\vec{S}_q$ is given by Eq.(\ref{equ-iii-2}). 
We want to treat this system for low temperatures. To this end we
introduce a parameter $\epsilon$ to be the inverse of the
spontaneous magnetization
\be
\la{equ-iii-47}
\epsilon = \left( \frac{\langle S^z_0 \rangle}{N} \right)^{-1} = 
\sigma^{-1} (T)\quad ,
\ee
and scale the time to
\be
\la{equ-iii-48}
\tau = \sigma t \quad ,
\ee
and the spin operators according to
\be
\la{equ-iii-49}
\hat{S}_q^\pm = \frac{S^\pm_q}{\sqrt{2\sigma}} \quad .
\ee
The operators $\delta S^z_q = S^z_q - \langle S^z_q \rangle$ will not 
be affected. Then
the equations of motion for the scaled spins read
\bea
\frac{d}{d\tau} \delta S^z_1 &=&  - \frac{2i}{N} \sum_2
(J_{1-2} - J_2) \hat{S}_2^{-} \hat{S}^+_{1-2} \la{equ-iii-50} \\
\frac{d}{d\tau}  \hat{S}^\pm_1 &=& \mp 2i (J_0 - J_1) 
\hat{S}^\pm_1 \pm \epsilon \frac{2i}{N} \sum_2 (J_{1-2} - J_2) 
\delta S^z_2 \hat{S}^\pm_{1-2} \la{equ-iii-51} \quad,
\eea
where we have used the shorthand notations $q_1 \rightarrow 1,
q_1-q_2 \rightarrow 1-2$. The parameter $\epsilon (T)$ will
not approach zero, except for the case $S \rightarrow \infty$, but one can
directly see from (\ref{equ-iii-50}) and (\ref{equ-iii-51}) that the
zeroth order contribution in $\delta S^z_1(\tau)$ and $S^\pm_1 (\tau)$ 
gives the correct spin wave approximation in lowest order 
\cite{8}\footnote{As referred to 
time $t$ we have the renormalized spin wave 
frequencies
$\omega_1(T) = 2 \sigma(T)(J_0-J_1)$.}, so that introducing $\epsilon$ will
allow for a low temperature approximation. This will be confirmed by the
final results.

We want to apply the theory of section 2 to the dynamics with respect 
to $\tau$, which
means that we formally have a Liouvillian $\tilde{L} = L/\sigma$,
or Hamiltonian $\tilde{{\cal H}} = {\cal H}/\sigma$, 
respectively. As a consequence we have a scaled temperature 
$\tilde{\beta}$ with
$\tilde{\beta} {\tilde {\cal H}} = \beta {\cal H}$. Choosing the set of 
observables
$G$ to be all modes $\dsz_q$ and $\shpm_q$
\be
\la{equ-iii-52}
\{G\} = \{\delta S^z_q, \hat{S}^\pm_q\} \quad ,
\ee
and inspecting the equations of motion (\ref{equ-iii-50}) 
and (\ref{equ-iii-51})
for $G$ one sees that it holds
\be
\la{equ-iii-53}
\tilde{L} G = V^{(1)} G + V^{(2)} GG
\ee
with
\be
\la{equ-iii-54}
V^{(2)} = V^{(2)}_0 + \epsilon V^{(2)}_1 \quad ,
\ee
so that our condition for an expansion of the residual forces is fulfilled.
Furthermore the $G, 1$ form the Lie-algebra
\bea
\left[ \hat{S}^{+}_1 , \hat{S}^{-}_2\right] &=& 
N \delta_{1,-2} + \epsilon \delta S^z_{1+2}
\la{equ-iii-55} \\
\left[ \delta S^z_1, \hat{S}^\pm_2\right] &=& \pm \hat{S}^\pm_{1+2}
\quad , \la{equ-iii-56}
\eea
which will be important for the statics.
\subsubsection{Residual forces and Heisenberg dynamics}
We start with the exact equations (\ref{equ-ii-17}) for the residual
forces which follow for our special dynamics. Regarding that the static
and dynamic correlation matrices are diagonal with respect to wave numbers
and components $\alpha = z, +, -,$ and introducing
\bea
\Xi^z_1 (\tau) & =&  \frac{(\delta S^z_1|\delta S^z_1(\tau))}
{(\delta S^z_1|\delta S^z_1)}
\la{equ-iii-57}\\
\Xi^\pm_1 (\tau) & = & \frac{(\hat{S}^\pm_1|\hat{S}^\pm_1(\tau))}
{(\hat{S}^\pm_1|\hat{S}^\pm_1)}
 =  \frac{(S_1^\pm|S_1^\pm(\tau))}{(S^\pm_1|S^\pm_1)}
\la{equ-iii-58}
\end{eqnarray}
we have\footnote{The convolutions refer to $\tau$.}
\begin{eqnarray}
\la{equ-iii-59}
f^z_1(\tau) & = & - \frac{2i}{N} \sum_2 (J_{1-2} - J_2) Q \left( 
\hat{S}^{-}_2 \Xi^{-}_2
+ \hat{f}_2^{-} \otimes  \Xi^{-}_2 \right) \left( \hat{S}^{+}_{1-2} 
\Xi^{+}_{1-2} 
+ \hat{f}_{1-2}^{+} \otimes \Xi^{+}_{1-2} \right)  \nonumber \\
& + & f^z_1 \otimes \frac{2i}{N} \sum_2 
\frac{(J_{1-2} - J_2) }{(\delta S^z_1|\delta S^z_1)}\nonumber\\
& & \cdot 
\left( \delta S^z_1|\left( \hat{S}^{-}_2 \Xi^{-}_2 + \hat{f}_2^{-} \otimes
\Xi^{-}_2 \right) 
\left( \hat{S}^{+}_{1-2} \Xi^{+}_{1-2} + \hat{f}_{1-2}^{+} 
\otimes \Xi^{+}_{1-2} \right) \right) \quad ,
\end{eqnarray}
and
\begin{eqnarray}
\la{equ-iii-60}
\hat{f}^\pm_1(\tau) & = & \pm \epsilon \frac{2i}{N} 
\sum_2 (J_{1-2} - J_2) Q \left( \delta S^z_2 \Xi^z_2
+ f_2^z \otimes \Xi^z_2 \right)
\left( \hat{S}^\pm_{1-2} \Xi^\pm_{1-2} 
+ \hat{f}_{1-2}^\pm \otimes \Xi^\pm_{1-2} \right)  \nonumber\\
& \mp & \epsilon \hat{f}^\pm_1 (\tau) \otimes \frac{2i}{N} 
\sum_2 
\frac{(J_{1-2} - J_2) }{(\hat{S}^\pm_1|\hat{S}^\pm_1)}  \nonumber \\
& & \cdot \left( S^\pm_1| \left( \delta S^z_2 \Xi^z_2 
+ f_2^z \otimes \Xi^z_2 \right) \left( \hat{S}^\pm_{1-2} \Xi^\pm_{1-2} 
+ \hat{f}_{1-2}^\pm \otimes \Xi^\pm_{1-2} \right) \right) \quad .
\end{eqnarray}
One directly sees that the transversal forces 
$\hat{f}^\pm_1 (\tau)$ start with a 
first order contribution in $\epsilon$ whereas the longitudinal forces
$f^z_1 (\tau)$ have a term of zeroth order. This structure of 
Eqs.(\ref{equ-iii-59})
and (\ref{equ-iii-60}) allows for a solution in powers of $\epsilon$.
Restricting to the non--vanishing lowest order approximation\footnote{For
$f^\pm_1 (\tau)$ the first order contribution is necessary as to obtain spin
wave damping, whereas the first order correction to $f^z_1 (\tau)$ would
just modify $\Xi^z_1(\tau)$ quantitatively.} we find
\bea
f^z_1 (\tau) &=& - \frac{2i}{N} \sum_2 \, \, Q \hat{S}^{-}_2 \, \, 
\hat{S}^{+}_{1-2}
(J_{1-2} - J_2) \phi_{12} (\Xi^{+}, \Xi^{-})
\la{equ-iii-61}\\
\hat{f}^{\pm}_1 (\tau) & = & \pm \epsilon \frac{2i}{N} \sum_2 (J_{1-2} - J_2) Q
\left\{ 
\vphantom{\frac{2i}{N}}
\delta S^z_2 \hat{S}^{\pm}_{1-2} \Xi^z_2 \Xi^{\pm}_{1-2}
\right. \nonumber \\
& & \left. - \frac{2i}{N} 
\sum_3 \, \, Q \left( \hat{S}^{-}_3 
\hat{S}^{+}_{2-3} \right)
\hat{S}^{\pm}_{1-2} (J_{2-3} - J_3) \left( {\phi}_{23} \otimes \Xi^z_2 \right)
\Xi^{\pm}_{1-2} \right\}  \la{equ-iii-62} \quad ,
\end{eqnarray}
where $\phi_{12}$ is a functional of the correlations $\Xi^{+}, \Xi^{-}$ and
can be expressed by Laplace transform to yield
\be
\la{equ-iii-63}
\phi_{12}(s) = \frac{\displaystyle
\left( \Xi^{-}_2 \Xi^{+}_{1-2} \right) (s)}
{\displaystyle
1 - \frac{2i}{N} \sum\limits_{3} \frac{\left( \delta S^z_1|\hat{S}^{-}_3 \hat{S}^{+}_{1-3} 
\right)} {\left( \delta S^z_1|\delta S^z_1 \right)} (J_{1-3} - J_3)
(\Xi_3^{-} \Xi^{+}_{1-3}) (s)} \quad .
\ee
The Heisenberg dynamics of the spins follow from the general relations 
(\ref{equ-ii-10}).
Regarding the symmetry of the correlation matrices we find
\bea
\delta S^z_1 (\tau) &=& \delta S^z_1 \Xi^z_1 (\tau) + f^z_1 (\tau) \otimes
\Xi^z_1 (\tau) \la{equ-iii-64}\\
\hat{S}^\pm_1 (\tau) &=& \hat{S}^\pm_1 \; \Xi^\pm_1 (\tau) +
\hat{f}^\pm_1 (\tau) \otimes \Xi^\pm_1 (\tau) \la{equ-iii-65} \quad .
\eea
Inserting the approximations for the residual forces (\ref{equ-iii-61}) and
(\ref{equ-iii-62}) into (\ref{equ-iii-64}) and (\ref{equ-iii-65}) respectively,
yields the corresponding approximations for the Heisenberg dynamics. One sees
that the longitudinal components $\delta S^z_1 (\tau)$ move in a 
subspace spanned
by $\delta S^{z}_1 $ and all products $\hat{S}^{-}_2 \hat{S}^{+}_{1-2}$,
whereas the transverse components are restricted to a space spanned by
$\hat{S}_1^\pm$ and all the products $\delta S^z_2 \hat{S}^\pm_{1-2}$ and
$\hat{S}^{-}_3 \hat{S}^{+}_{2-3} \hat{S}^{+}_{1-2}$ respectively
\bea
\delta S^z_1 (\tau) &=& \cdots  \delta S^z_1 + \sum_2 \cdots \hat{S}^{-}_2
\hat{S}^{+}_{1-2}
\la{equ-iii-66}\\
\hat{S}^\pm_1 (\tau) &=& \cdots \hat{S}^\pm_1 + \epsilon \sum_2 \cdots 
\delta S^z_2 \hat{S}^\pm_{1-2} + \epsilon \sum_{2,3} \cdots \hat{S} ^{-}_3
\hat{S}^{+}_{2-3} \hat{S}^\pm_{1-2}
\la{equ-iii-67} \quad .
\eea
The time dependent coefficients are given by functionals of the dynamic
correlations $\Xi^z (\tau), \Xi^\pm (\tau)$, and the exchange parameters, and
static correlations. It should be noted that the results for 
$\delta S^z (\tau)$
and $\hat{S}^\pm (\tau)$ have the same subspaces as would have been
obtained by a simple perturbational expansion of the dynamic equations
(\ref{equ-iii-50}) and (\ref{equ-iii-51}). The essential difference of 
(\ref{equ-iii-66}) and (\ref{equ-iii-67}) to such an expansion is, that
the coefficients of $\delta S^z_1$ and $\hat{S}^\pm_1$ in (\ref{equ-iii-66})
and (\ref{equ-iii-67}) comprise all contributions of the simple perturbational
series which have a projection onto $\delta S^z, \hat{S}^\pm$. As compared to
a Holstein Primakoff approach\footnote{Expanding the coefficients in 
(\ref{equ-iii-67}) into powers of $\epsilon$ with $\epsilon (T=0) = 1/S$ and
inserting the expansions of $\hat{S}^\pm$ into Bose operators 
$a, a^{\dagger}$ into
(\ref{equ-iii-67}), then up to order $S^{-1}$ the result for 
$\hat{S}^{+}_1(\tau)$ coincides with the dynamics of $a(\tau)$ 
calculated with the four
magnon interaction as a perturbation.}, $\hat{S}^\pm (\tau)$ corresponds to
a perturbational treatment of the four magnon interaction, if in addition,
the Bose operators are summed up to give the spin operators, and the
coefficients will be renormalized.
\subsubsection{Dynamic and static correlations}
The dynamics of the correlation matrix $\Xi$ is governed by the Mori
equations (\ref{equ-ii-9}). Regarding the symmetry of the Hamiltonian we get
\bea
\frac{d}{d\tau} \Xi^z_1 &=& - \gamma_1^{\parallel} \otimes \Xi^z_1
\la{equ-iii-68} \\
\frac{d}{d\tau} \Xi^\pm_1 &=& \mp i \omega_1 \Xi^\pm_1 - \gamma^{\bot}_1 
\otimes \Xi^\pm_1 \quad ,\la{equ-iii-69}
\eea
where we have introduced the memory functions
\bea
\gamma_1^{\parallel} (\tau)  &=&  \frac{(f^z_1|f^z_1(\tau)) }
{(\delta S^z_1|\delta S^z_1)}
=  \frac{-i}{\tilde{\beta}(\delta S^z_1|\delta S^z_1)}  \langle 
[S^z_{-1}, f^z_1(\tau)]\rangle 
\la{equ-iii-70}\\
\gamma^\bot_1 (\tau)  &=&  \frac{(\hat{f}^{+}_1|\hat{f}^{+}_1 (\tau)) }
{(\hat{S}^{+}_1|\hat{S}^{+}_1)} 
 =  \frac{-i}{\tilde{\beta}(\hat{S}^{+}_1|\hat{S}^{+}_1)} 
\langle [\hat{S}^{-}_{-1}, \hat{f}^{+}_1 (\tau)] \rangle \quad ,
\la{equ-iii-71}
\eea
and the frequency
\be
\la{equ-iii-72}
\omega_1(T) = N \left( \tilde{\beta} (\hat{S}^{+}_1|\hat{S}^{+}_1) 
\right)^{-1} .
\ee
In the following we will use the commutator forms for $\gamma^{\parallel}$
and $\gamma^{\bot}$\footnote{We avoid Mori products of the form $(GG|A)$.}.
Inserting the approximations for $f^z_1 (\tau)$ and $f^\pm_1 (\tau)$
(\ref{equ-iii-61}), (\ref{equ-iii-62}) into (\ref{equ-iii-70}) and
(\ref{equ-iii-71}), we obtain the following expressions for the memory
functions:
\bea
\gamma_1^{\parallel}(\tau) &=& \frac{2}{N} \sum_2 (J_{1-2} - J_2) A_{12}
(T) \phi_{12} [\Xi^{+} , \Xi^{-}]
\la{equ-iii-73}\\
\gamma^\bot_1 (\tau) & = & \epsilon \frac{2}{N} \sum_2 (J_{1-2} - J_2) B_{12}
(T) \Xi^z_2 \Xi^{+}_{1-2}  \nonumber \\
& + &  \epsilon \left( \frac{2}{N} \right)^2 \sum_{2,3} (J_{1-2} - J_2)
(J_{2-3} - J_3) C_{12,3} (T) \left( \phi_{23} \otimes \Xi^z_2 \right) 
\Xi^{+}_{1-2} \quad ,
\la{equ-iii-74}
\end{eqnarray}
where the temperature dependent coefficients are given by
\bea
A_{12} (T) &=& \frac{1}{\tilde{\beta}(\delta S^z_1|\delta S^z_1)}
\left( \langle \hat{S}^{-}_{2-1} \hat{S}^{+}_{1-2}\rangle - 
\langle \hat{S}^{-}_2 \hat{S}^{+}_{-2} \rangle \right)
\la{equ-iii-75}\\
B_{12} (T) &=& \frac{1}{\tilde{\beta}(\hat{S}^{+}_1|\hat{S}^{+}_1)}
\left( \langle \hat{S}^{-}_{2-1} \hat{S}^{+}_{1-2}\rangle - \epsilon 
\langle \delta S^z_2
\delta S^z_{-2} \rangle - \frac{(\hat{S}^{+}_1|\delta S^z_2 
\hat{S}^{+}_{1-2})}
{(\hat{S}^{+}_1|\hat{S}^{+}_1)} \right )
\la{equ-iii-76}\\
C_{12,3} (T) & = & \frac{(\delta S^z_2|\hat{S}^{-}_3 \hat{S}^{+}_{2-3})}
{(\delta S^z_2|\delta S^z_2)} B_{12} (T)\nonumber\\ 
 &+&  \frac{1}{\tilde{\beta}(\hat{S}^{+}_1|\hat{S}^{+}_1)}
\left( (\delta_{3,2-1} + \delta_{2,0}) N \langle \hat{S}^{-}_3 
\hat{S}^{+}_{-3}\rangle - 
\frac{(\hat{S}^{+}_1|\hat{S}^{-}_3 \hat{S}^{+}_{2-3} \hat{S}^{+}_{1-2})}
{(\hat{S}^{+}_1|\hat{S}^{+}_1)} 
\right.\nonumber\\
& & \left. + 
\epsilon \langle  \hat{S}^{-}_3 \delta S^z_{2-3} 
\hat{S}^{+}_{1-2}\rangle + \epsilon
\langle \hat{S}^{-}_3 \hat{S}^{+}_{2-3} \delta S^z_2 \rangle 
\vphantom{
\frac{(\hat{S}^{+}_1|\hat{S}^{-}_3 \hat{S}^{+}_{2-3} \hat{S}^{+}_{1-2})}
{(\hat{S}^{+}_1|\hat{S}^{+}_1)} }
\right) \quad .
\la{equ-iii-77}
\end{eqnarray}
Inspecting the time dependence of the memory function, one sees that 
$\gamma^{\parallel}$ 
and $\gamma^{\bot}$ are determined by $\Xi^{+}, \Xi^{-}$, or 
$\Xi^{+}, \Xi^{-}, \Xi^z$, respectively. Thus we have a nonlinear coupled
set of equations for $\Xi^z, \Xi^{+}, \Xi^{-}$. 
This set of equations is very complicated, as it is nonlinear
and contains static correlations which must be known, or determined from
the dynamic equations. For this reason we first state, how the static 
correlations
in principle can be obtained in a selfconsistent way. Then we 
calculate their leading terms for $T \rightarrow 0$ and show, how
the equations for $\Xi^z$ and $\Xi^{+}$ reduce to the standard forms.

The static correlations entering into $\Xi^z$   and $\Xi^\pm$ are found
from Eqs. (\ref{equ-iii-75})--(\ref{equ-iii-77}) to read
\begin{equation}
\la{equ-iii-79}
\begin{array}{cccc}
(\delta S^z|\delta S^z), &  \langle \delta S^z \delta S^z \rangle,&
(\hat{S}^{+}|\hat{S}^{+}), & \langle \hat{S}^{-} 
\hat{S}^{+} \rangle, \\
(\delta S^z | \hat{S}^{-} \hat{S}^{+}), & (\hat{S}^{+} |\delta S^z 
\hat{S}^{+}), &
(\hat{S}^{+}|\hat{S}^{-} \hat{S}^{+} \hat{S}^+), & \langle 
\hat{S}^{-} \hat{S}^{+} \delta \hat{S}^z\rangle
\end{array} .
\end{equation}
If additionally one takes into account the correlations
\be
\la{equ-iii-80}
\langle \delta \hat{S}^z \delta \hat{S}^z \rangle , \quad
\langle \hat{S}^{-} \hat{S}^{-} \hat{S}^{+} \hat{S}^{+}\rangle, \quad
\langle \hat{S}^{-} \hat{S}^{+} \delta \hat{S}^z \delta \hat{S}^z \rangle,
\quad 
\langle \hat{S}^{-} \hat{S}^{-} \hat{S}^{+} 
\hat{S}^{+} \delta \hat{S}^z \rangle \quad ,
\ee 
then one can find a closed set of equations for this extended set
(\ref{equ-iii-79}), (\ref{equ-iii-80}). The procedure is similar to the
considerations in section 2.5. The difference is, that our iteration with
$V^{(2)}_0 + \epsilon V^{(2)}_1$ has lead  to residual forces 
which also have contributions
of the form $GGG$, so that the commutators are modified. The details are 
given in appendix C.

For our purpose, however, it is sufficient to have the static correlations
for $T \rightarrow 0$. To this end we iterate the system of static equations
with respect to $\epsilon$, as the first order contributions result to be
smaller than those of zeroth order. In zeroth order the coupled system of
dynamic and static correlations for $\Xi^z_1, \Xi^\pm_1$ and the correlations
(\ref{equ-iii-79}) can be solved exactly. Then the expressions of first order
follow in a straight forward manner. As those terms we need are exact, they
can also be obtained in a different way. Introducing
\be
\la{equ-iii-81}
\tilde{\omega}_1 = 2 (J_0 - J_1)
\ee
\be
\la{equ-iii-82}
n_1 = \frac{1}{e^{\tilde{\beta} \tilde{\omega}_1} - 1}
\ee
we explicitly find for the transverse correlations\footnote{The results
are given in zeroth order, and up to first order, where they will be needed.}
\bea
\langle \hat{S}^{-}_{-1} \hat{S}^{+}_1 \rangle &=& N n_1 + \epsilon n_1
(1 + n_1) \tilde{\beta} \sum_2 (\tilde{\omega_2} - \tilde{\omega}_{1-2}) n_2 
+ {\cal O}(\epsilon^2)
\la{equ-iii-83}\\
\frac{(\hat{S}^{+}_1|\delta S^z_2 \hat{S}^{+}_{1-2})}{(\hat{S}^{+}_1|
\hat{S}^{+}_1)}
& = & - n_{1-2}  \nonumber \\
& - & \epsilon \left\{ n_{1-2} (1+n_{1-2}) \frac{\tilde{\beta}}{N} \sum_3
(\tilde{\omega}_3 - \tilde{\omega}_{1-3}) n_3 
-  \frac{1}{N} 
\sum_3 n_3 (1 + n_{2-3})  \right.  \nonumber \\ 
&+ & \left.   \frac{1}{N} \sum_3 [n_3 (1+n_{1-2} + n_{2-3}) - 
n_{1-2} n_{2-3} ]  \right. \nonumber \\
& & \left . \cdot 
\frac{\tilde{\omega}_2 - \tilde{\omega}_{1-2} + \tilde{\omega}_{1+3-2} 
- \tilde{\omega}_{2-3}}{\tilde{\omega}_3 - \tilde{\omega}_{1-2} 
- \tilde{\omega}_{2-3}} \right \}  + {\cal O}(\epsilon^2)
\la{equ-iii-84}\\
\frac{(\hat{S}^{+}_1|\hat{S}^{-}_3 \hat{S}^{+}_{2-3} \hat{S}^{+}_{1-2})}
{(\hat{S}^{+}_1|\hat{S}^{+}_1)}
& = & N (\delta_{3,2-1} + \delta_{2,0}) n_3  \nonumber \\
& + & \epsilon \left\{ 
\vphantom{\frac{\tilde{\omega}_2 - \tilde{\omega}_{1-2} + 
\tilde{\omega}_{1+3-2}
- \tilde{\omega}_{2-3}} {\tilde{\omega}_3 - \tilde{\omega}_{1-2}
- \tilde{\omega}_{2-3}} }
(\delta_{3,2-1} + \delta_{2,0}) n_3 (1+n_3)
\tilde{\beta} \right .\nonumber\\
& & \cdot
\sum_4 (\tilde{\omega}_4 - \tilde{\omega}_{3-4})n_4 -
n_3 (1 + n_{1-2} + n_{2-3})  \nonumber \\
& + &  [n_3  (1 + n_{1-2} + n_{2-3}) - n_{1-2} n_{2-3} ] 
 \nonumber \\
& & \left. \cdot  
\frac{\tilde{\omega}_2 - \tilde{\omega}_{1-2} + 
\tilde{\omega}_{1+3-2}
- \tilde{\omega}_{2-3}} {\tilde{\omega}_3 - \tilde{\omega}_{1-2}
- \tilde{\omega}_{2-3}} 
\right\} + {\cal O}  (\epsilon^2) ,
\la{equ-iii-85}
\end{eqnarray}
whereas the longitudinal correlations are calculated to yield
\bea
\langle \delta S^z_{-1} \delta S^z_{1} \rangle &=& \sum_2 n_2 (1 + n_{1-2})
+ {\cal O} (\epsilon)\la{equ-iii-86}\\
\langle \hat{S}^{-}_{-2} \hat{S}^{+}_{2-1} \delta S^z_1 \rangle &=& - N n_2
(1 + n_{1-2}) + {\cal O} (\epsilon)\la{equ-iii-87}\\
(\delta S^z_1|\delta S^z_1) &=& \sum_2 \frac{n_2 - n_{1-2}}
{\hat{\beta} (\tilde{\omega}_{1-2} - \tilde{\omega}_2)} + {\cal O} (\epsilon)
\la{equ-iii-88}\\
(\hat{S}^{-}_{1-2} \hat{S}^{+}_2|\delta S^z_1) &=& - N \frac{n_2 - n_{1-2}}
{\tilde{\beta} (\tilde{\omega}_{1-2} - \tilde{\omega}_2)} + {\cal O} (\epsilon)
\la{equ-iii-89} \quad .
\eea
All the static correlations (\ref{equ-iii-83})--(\ref{equ-iii-89}) listed 
above still depend on the spontaneous magnetization 
$\sigma (T) = \epsilon^{-1}$.
This can be determined from the condition 
$\sum_i \vec{s}_i \cdot \vec{s}_i = N  S(S+1)$ 
which can be cast into the form
\bea
\la{equ-iii-90}
0 &=&
(\sigma - S) + (\sigma - S) \left( \frac{2}{N^2} \sum_1 
\langle \hat{S}^{-}_{-1}
\hat{S}^{+}_1 \rangle (\sigma) + 2S + 1 \right)\nonumber\\ 
&+& \frac{1}{N^2} \sum_1
\left( \langle \delta S^z_{-1} \delta S^z_1 \rangle (\sigma) + 2 S
\langle \hat{S}^{-}_{-1} \hat{S}^{+}_1 \rangle (\sigma) \right)
\eea
and leads to the magnon result
\be
\la{equ-iii-91}
\sigma = S - \frac{1}{N} \sum\limits_{1} n_1  + \cdots \quad .
\ee
Once the statics are known we now can go back to the dynamics of $\Xi^z$
and $\Xi^{\pm}$, calculating the lowest order contribution to the memory
functions (\ref{equ-iii-73}) and (\ref{equ-iii-74}). For 
$\gamma_1^{\parallel} (s)$ 
we find the zeroth order result
\bea
\la{equ-iii-92}
\gamma_1^{\parallel}(s) 
&=& \frac{1}{s}  \left(\sum_{2} \frac{n_{1-2} - n_2}
{(\tilde{\omega}_2 - \tilde{\omega}_{1-2})\left(s-i(\tilde{\omega}_2 -
\tilde{\omega}_{1-2})\right)} \right)^{-1}\nonumber\\
& & \cdot
\sum_2 \frac{(\tilde{\omega}_2 - \tilde{\omega}_{1-2})
(n_{1-2} - n_2)}{s-i(\tilde{\omega}_2 - \tilde{\omega}_{1-2})}
\eea
which leads to
\be
\la{equ-iii-93}
\Xi_1^z (\tau) = \left( \sum_{2}\frac{n_{1-2} - n_2}{\tilde{\omega}_2 - 
\tilde{\omega}_{1-2}}\right)^{-1}
\sum_{2} \frac{n_{1-2} - n_2}{\tilde{\omega}_2 - \tilde{\omega}_{1-2}}
e^{i(\tilde{\omega}_2 - \tilde{\omega}_{1-2})\tau} \quad ,
\ee
whereas the memory function $\gamma_{\bot} (\tau)$ is of second order and
reads
\begin{eqnarray}
\la{equ-iii-94}
\gamma^{\bot}_1(\tau) & = & (\frac{\epsilon}{N})^2 \tilde{\beta} 
\tilde{\omega}_1
\sum_{2,3} (\tilde{\omega}_2 - \tilde{\omega}_{1-2})
\frac{\tilde{\omega}_2 - \tilde{\omega}_{1-2} + \tilde{\omega}_{1+3-2} -
\tilde{\omega}_3}{\tilde{\omega}_3 - \tilde{\omega}_{2-3} - 
\tilde{\omega}_{1-2}} \nonumber \\
& &\cdot   [n_{1-2} n_{2-3} (1+n_3) - (1+n_{1-2})(1+n_{2-3}) n_3]
\mbox{e}^{i(\tilde{\omega}_3 - \tilde{\omega}_{2-3} - 
\tilde{\omega}_{1-2})\tau} \quad .
\end{eqnarray}
The spin wave frequency (\ref{equ-iii-72}) up to second order yields
\begin{eqnarray}
\la{equ-iii-95}
\omega_1(T) & = & \tilde{\omega}_1
- \frac{\epsilon}{N} \sum_2 (\tilde{\omega}_2 - \tilde{\omega}_{1-2}) n_2 
\nonumber \\
& - & (\frac{\epsilon}{N})^2 \sum_{2} (\tilde{\omega}_2 - \tilde{\omega}_{1-2})
\sum_{3} n_3 [n_{2-3} + n_2 (1+n_2) \tilde{\beta} (\tilde{\omega}_3 - 
\tilde{\omega}_{1-3})] \nonumber \\
& & - \frac{\gamma^\bot_1(\tau=0)}{\tilde{\beta}\tilde{\omega}_1} \quad .
\end{eqnarray}
Comparing our results for $T \rightarrow 0$ to those of spin wave theory
and regarding
\bea
\tau  \tilde{\omega}_1 &=& t \cdot 2 \sigma (T) (J_0 - J_1)  \\
\tilde{\beta}  \tilde{\omega}_1 &=& \beta \cdot 2 \sigma (T)
(J_0 - J_1)
\eea
we see that $\Xi^z(t)$ coincides with the correlation function obtained by
Lovesey \cite{8}, 
provided we take our temperature dependent spin wave frequency
$2\sigma (T) (J_0 - J_1)$ at $T=0$. The memory function $\gamma^\bot_1 (t) = 
\gamma^\bot_1 (\tau) (d\tau/dt)^2$ (\ref{equ-iii-94}) agrees with 
the Holstein Primakoff result for the four magnon interaction \cite{5}, 
if again
$\sigma(T) \rightarrow S$ is used.

It is clear that the presented procedure for the Heisenberg ferromagnet is
rather complicated. Just for calculating spin wave frequencies and dampings 
one would choose the usual way. But we think it is important to have shown
that one is not forced to expand spin operators into Bose operators, as to
obtain the correlation functions at low temperatures. In this context our
main result is, that the Heisenberg operators $\vec{s}_i(t)$ 
(c.f. Eqs.(\ref{equ-iii-64}) and (\ref{equ-iii-65}))
can be expressed in terms of 
correlation functions and spin operators at $t=0$. This point will be
useful, if one wants to derive equations of motion for expectation values
of spins which go beyond linear response, as in such a case the expectation
value of $\langle S_q^{+} \rangle (t)$ cannot simply be replaced by the
expectation value of one Bose operator $\langle a_q \rangle (t)$
\section{Conclusion}
It has been shown, that incorporating the time evolution of products of
observables $(\g\g)(t) = \g(t)\g(t)$ into the frame work of Mori's theory,
one can give selfconsistent approximations for the residual forces. These
can be expressed in terms of time--dependent 
correlation functions and operators
at $t=0$. In this way it is possible to deduce approximations for the 
Heisenberg
dynamics of the observables as well as for the dynamic and static correlation
functions. We have tested this approach comparing the approximations to the
exact solution of a  model and to the theory of interacting spin waves. From
this comparison we conclude that the selfconsistent approximations in
Mori's theory can be successfully used to attack the dynamics of correlation
functions and Heisenberg operators of interacting systems.

The dynamic equations for the correlation functions we have found, are of
the type of mode--mode coupling equations, but the coefficients involved are
given by bare interactions. In addition, our approach allows for a quantitative
estimation of the validity of the presented
mode--mode coupling approximation. A point
of further investigation would be, how the expansion scheme can be generalized
to give mode--mode coupling equations with the bare interactions replaced by
matrix elements of the Liouville operator.

In our illustrations we have been concerned with spin systems. But the
given formalism also applies to Fermi or Bose systems, if for the
statics of Fermion systems the fluctuation--dissipation theorems
are used in an adequate way. Thus, the possibility
arises to relate expansions in Mori's theory to other many-body treatments,
and especially find a link between projection--operator expansions and those
in Green's function theories.

Summarizing we think that the main advantage of the presented expansions is,
that they lead to approximations for the Heisenberg dynamics. These can be
used to derive equations of motion for expectation values which go beyond
linear response.
\begin{appendix}
\section{Second order approximations to the residual forces}
\subsection{Expansion into powers of $\epsilon \sim V^{(2)}$ with fixed
correlations $\Xi(t)$}

The expansion of the residual forces in powers of $\epsilon$, with fixed
dynamic and static correlations, is obtained from (\ref{equ-ii-17}). 
The first
order result was given in (\ref{equ-ii-20}). Up to second order one finds
\begin{eqnarray}
\la{equ-A-1}
f(t) & = & i Q V^{(2)} \{\gx, \gx \}  \nonumber \\
& + & i Q V^{(2)} \{ i Q V^{(2)} \{\gx, \gx\} 
\otimes \Xi, \gx \}
 +  i Q V^{(2)} \{ \gx , i Q V^{(2)} \{\gx, \gx\} 
\otimes \Xi \} \nonumber \\
& - &  i Q V^{(2)} \{\gx, \gx \} \otimes (\g|\g)^{-1} 
(\g| i V^{(2)}\{\gx, \gx\}) + {\cal O} (\epsilon^3) \quad .
\end{eqnarray}
The approximate $f(t)$ now is an element of a linear space spanned by the
products $Q \g \g$ and $Q \g \g \g$. The selfconsistent equations for the
dynamic correlations $\Xi(t)$ are given by Eq.(\ref{equ-ii-9}) with the force
correlation functions $(f|f(t))$ calculated from (\ref{equ-A-1}). 
A general feature
of the expansion (\ref{equ-A-1}) 
of $f(t)$ is that terms of second, and higher order,
will contribute to the projections onto the forces at $t=0$
\be
\la{equ-A-2}
 f = i Q V^{(2)}\{\g, \g\} \quad .
\ee
That means they will contribute to the force correlation functions. 
In the sense 
of an orthogonal decomposition of $f(t)$, the coefficients of $Q \g \g$ are
changed by each order of the iteration. This is, why it seems to be reasonable,
to sum up all contributions of higher order terms which are parallel to the
products $Q \g \g$. To make this reasoning explicit, we introduce a basis
$F_{\nu}$ for the $Q \g \g$ by
\be
\la{equ-A-3}
F_{\nu} = Q \sum_{\mu, \lambda} \; 
\alpha_{\nu, \mu \lambda} (\g_\mu \g_\lambda -
\langle \g_\mu \g_\lambda \rangle)
\ee
which is abbreviated by
\be
\la{equ-A-4}
F= Q \alpha \{\g, \g \} - \langle \alpha \{\g, \g \}\rangle \quad .
\ee
Then $f(t)$ can be decomposed as
\be
\la{equ-A-5}
f(t) = F (F|F)^{-1} (F|f(t)) + \qt f(t) \quad ,
\ee
where
\be
\la{equ-A-6}
\qt = \{(1-|F)(F|F)^{-1} (F|\} Q \quad ,
\ee
and $(F|f(t))$ is no longer affected by the approximations of $\qt f(t)$
which are orthogonal to (\ref{equ-A-3}). 
In the next section A.2 we will calculate $f(t)$
to second order with fixed $\Xi(t)$ and fixed components $(F|f(t))$.
\subsection{Expansion into powers of $\epsilon \sim V^{(2)}$ with fixed
correlations $\Xi(t)$ $and$ $\Gamma (t)$}
For a treatment of $f(t)$ according to (\ref{equ-A-5}) 
it is expedient to decompose 
$f(t)$ into a sum 
\be
\la{equ-A-7}
f(t) = e^{iQLQt} iQL\g = K(t) i \Omega^{(r)} \quad ,
\ee
where
\be
\la{equ-A-8}
K(t) = e^{iQLQt} \f \quad ,
\ee
and the matrix $\Omega^{(r)}$ is defined by
\be
\la{equ-A-9}
\Omega^{(r)}  =  (\f|\f)^{-1} (\f|L\g) 
 =  \ff (\f |\vv \{ \g, \g \}) \quad .
\ee
Here use had been made of the fact that\footnote{From $(1|L\g)=0$ and
$(1|\g ) = 0 $ it follows $(1|\vv \{ \g, \g \}) = 0$.}
\be
\la{equ-A-10}
QL\g = Q \left( \v \g + \vv \{ \g, \g \} \right) = \f \Omega^{(r)} \quad .
\ee
Then corresponding to (\ref{equ-A-5}) we can write
\be
\la{equ-A-11}
K(t) = \f  \Gamma(t) + \qt  K(t)
\ee
with
\be
\la{equ-A-12}
\Gamma(t) = \ff (\f |K(t)) = \ff (\f |e^{i Q L Q t } \f ) \quad .
\ee
In the sequel we will focus on the expansion of the quantity $K(t)$, which
is of course equivalent to the expansion of the residual force $f(t)$.

For $K(t)$ we will set up an exact equation with fixed $\Xi(t)$ and
$\Gamma(t)$ which can be iterated. As a first step we cast $K(t)$ into the
form
\be
\la{equ-A-13}
K(t) = \f  \Gamma(t)  +  \qt e^{iLt} QiL\f  \otimes 
\Gamma(t) 
  -  i\qt e^{iLt} \f  \otimes \ff (L\f |K(t))
\ee
which can be verified by Laplace transform and the completeness relation
\be
\la{equ-A-14}
\f \ff (\f| Q \g \g ) 
= Q \g \g - \langle \g \g \rangle \quad .
\ee
Next we take the Heisenberg dynamics of $ \g $ and $\f $, 
which can be expressed
as
\bea
\g (t)  &=&  \gx (t) + K(t) i \Omega^{(r)} \otimes \Xi(t) \la{equ-A-15}\\
\f (t)  &=&  \gx (t) \otimes i \Omega^{(l)} \Gamma(t) + K(t)
\left( 1-\otimes \Omega^{(r)} \Xi(t) \Omega^{(l)} \otimes \Gamma(t) \right)
\la{equ-A-16}
\quad .
\eea
Here $ \Omega^{(l)}$ denotes the matrix
\be
\la{equ-A-17}
\Omega^{(l)} = (\g |\g )^{-1} (\g |L \f )
= (G|G)^{-1} (V^{(2)}\{G,G\}|F) \quad ,
\ee
and use has been made of Eqs.(\ref{equ-ii-9}) and (\ref{equ-A-7}), and
\be
\la{equ-A-18}
\dot{\g} = \f (t) i\Omega^{(r)} + \g (t) i\Omega = \g \dot{\Xi}(t) 
+ f(t) + f(t) \otimes \dot{\Xi}(t) \quad . 
\ee
Calculating $L\f$ from the definition (\ref{equ-A-3}) 
and inserting Eqs.(\ref{equ-A-15})
and (\ref{equ-A-16}) into the right hand side of Eq.(\ref{equ-A-13}) one 
arrives at an exact system of nonlinear equations for $K(t)$
\bea
\la{equ-A-19}
K(t)&=& F \Gamma(t)\\
&+& \tilde{Q}\alpha_s \left\{
- G \Xi(t) \otimes \Omega^{(l)} \Gamma(t) \Omega^{(r)} +
K(t) i \Omega^{(r)} \left( 1 - \otimes \Xi(t) \otimes \Omega^{(l)}
\Gamma(t) \Omega^{(r)} \right), \right . \nonumber\\
& & \left.
G \Xi(t) + K(t) i \Omega^{(r)} \otimes \Xi(t) \right\} \otimes \Gamma(t)
\nonumber\\
&-&
\tilde{Q} K(t) i \Omega^{(r)} \otimes \Xi(t) (G|G)^{-1} (G|\alpha_s
\{F i \Omega^{(r)},G\})\otimes \Gamma(t) \nonumber\\
&+&
\tilde{Q} K(t) \left(1-\otimes \Omega^{(r)} \Xi(t) \Omega^{(l)}
\otimes \Gamma(t) \right) \otimes (F|F)^{-1} \left(\alpha_s
\{F i \Omega^{(r)},G\}| K(t) \right) \nonumber
\eea
where the abbreviation
\be
\la{equ-A-19a}
\alpha_s \{A,B\}:= \alpha\{A,B\}+\alpha\{B,A\}
\ee
has been introduced.
This system corresponds to Eq.(\ref{equ-ii-17}) for $f(t)$. The difference is
that in Eq.(\ref{equ-A-19}) 
the correlation functions $\Xi(t)$ of the
observables {\it and} the correlations $\Gamma(t)$ 
(\ref{equ-A-12}) of the residual
forces (\ref{equ-A-8}) appear.

The system (\ref{equ-A-19}) now can be iterated. Remembering that according
to (\ref{equ-A-9}) and (\ref{equ-A-17}) $\Omega^{(r)}$ and $\Omega^{(l)}$
are of first order in $\vv $, we find from (\ref{equ-A-19}) for fixed
$\Xi(t), \Gamma(t)$ (in treating $\Gamma(t)$ fixed the summation of higher
order contributions is performed):
\be
\la{equ-A-20}
K(t) = \f \Gamma(t)  +  \qt \alpha_s \left\{ \f \Gamma(t) i \Omega^{(r)},
\gx (t) \right\} \otimes \Gamma(t) + {\cal O}(\epsilon^2) \quad .
\ee
According to Eq.(\ref{equ-A-7}) it is a second order result for $f(t)$.
The main point is, that the approximation (\ref{equ-A-20}) conserves the
exact relation
\bea
\la{equ-A-21}
(\g |\g )^{-1} (f|f(t))   &=&  (\g | \g )^{-1} (iQL\g |\f ) (\f |\f )^{-1}
(\f |K(t)) i \Omega^{(r)} \nonumber\\
& =&  \Omega^{(l)} \Gamma(t) \Omega^{(r)} \quad .
\eea
The matrix $\Gamma(t)$ can be calculated from the differential equation
\be
\la{equa-A-22}
\dot{\Gamma}(t) = i\ff (F|L\f ) \Gamma(t) - \ff (iL\f |\qt K(t))
\ee
which follows from Eqs.(\ref{equ-A-11}) and (\ref{equ-A-12}).
Inserting the approximation (\ref{equ-A-20}) for $\qt K(t)$ and using the
definition (\ref{equ-A-4}) we find
\begin{eqnarray}
\la{equ-A-23}
\dot{\Gamma}(t) &=& i\ff (F|L\f ) \Gamma(t)\nonumber\\
 &-& \ff 
\left( \left. \alpha_s \left\{ \f i \Omega^{(r)}, \g \right\}  
\right| \qt \left( \alpha_s \left\{\f \Gamma(t)
i\Omega^{(r)}, \gx (t) \right\}
\right)\right) \otimes \Gamma(t) \quad .
\end{eqnarray}
Together with the dynamic equation for $\Xi(t)$
\be
\la{equ-A-24}
\dot{\Xi}(t) = \Xi(t) i \Omega - 
\Xi(t) \otimes \Omega^{(l)} \Gt \oo 
\ee
we have a coupled system for the correlations $\Gt $ and $\Xi(t)$. As
Eq.(\ref{equ-A-24}) can be solved by Laplace transform in terms of
$\Gamma(s)$, Eq.(\ref{equ-A-23}) can be viewed as a nonlinear integral
equation for the Laplace transforms $\Gamma(s)$. It should be noted that
the approximations (\ref{equ-A-23}), (\ref{equ-A-24}) have the exact second
order moment of $\Gt $ or the exact forth moment for $\Xi(t)$ respectively.
For later use we note the Heisenberg dynamics for $\g (t)$ which result from
the approximation (\ref{equ-A-20}). One obtains
\begin{eqnarray}
\la{equ-A-25}
\g (t) &=& \gx (t)  +  f(t) \otimes \Xi(t) = 
\gx (t)  +  \f \Gt i \oo \otimes \Xi(t) \nonumber \\
&+& \qt \alpha_s \left\{ \f \Gt i \oo, \gx (t)\right\}  
\otimes \Gt i \oo \otimes \Xi(t) \quad .
\end{eqnarray}
\subsection{Expansion of the residual forces $\tilde{f}(t)$}
We want to show that the results of section A.2 can be obtained in a simple way
by extending the set of observables $\g $ to $\gt $ and applying the procedure
of section 2.2 to $\tilde{f}$. Let us choose a set $\gt $ by
\be
\la{equ-A-26}
\gt = \{\g_{\nu}, \f_{\mu}\}
\ee
where ${\f}_{\nu}$ is defined by (\ref{equ-A-3}). Then we have a projection 
operator (\ref{equ-A-6})
\be
\la{equ-A-27}
\qt = 1 - \tilde{P} = 1 - |\g) (\g |\g )^{-1} (\g |-| \f )\ff (\f |
\quad .
\ee
The matrix $\tilde{\Xi}(t)$ can be written in block matrices as
\be
\la{equ-A-28}
\tilde{\Xi} = \left( \begin{array}{lr}
\Xi_{GG} & \Xi_{GF} \\
\Xi_{FG} & \Xi_{FF} \end{array} \right) \quad .
\ee
Similar the frequency matrix $\tilde{\Omega}$ reads
\be
\la{equ-A-29}
\tilde{\Omega} = \left( \begin{array}{lr}
\Omega_{GG} & \Omega_{GF} \\ \Omega_{FG} & \Omega_{FF} \end{array} \right)
\ee
where $\Omega_{GG}=\Omega$, 
and  $\Omega_{GF}$ and $\Omega_{FG}$ coincide with
the definitions (\ref{equ-A-17}) and (\ref{equ-A-9}) for $\Omega^{(l)}$
and $\oo$.

First we study the derivatives $L\gt $ and prove that the 
conditions (\ref{equ-ii-14}) of section 2.2 apply.
The derivative
\be
\la{equ-A-30}
L\g = \v \g + \vv \{\g , \g \}
\ee
is a linear combination of $\g_{\nu}$ and $\f_{\nu}$. Hence in abbreviated
notation we have
\be
\la{equ-A-31}
L\g = \g \Omega_{GG} + \f \Omega_{FG} \quad .
\ee
The derivatives $L\f$ are found from (\ref{equ-A-4}). Together with
(\ref{equ-A-31}) one yields\footnote{We take the case that for symmetry
 reasons the constant term in (\ref{equ-A-32}) vanishes. If it does not
vanish it can be handled without difficulty.}
\begin{eqnarray}
\la{equ-A-32}
LF & = & \g \left[ (\g |\g )^{-1}(\g |\alpha_s \{\g \Omega_{GG}, \g \} 
) - \Omega_{GG} (\g |\g )^{-1} (\g |\alpha \{ \g ,\g \})\right]  \nonumber \\
& + & \f \left[ \ff (\f | \alpha_s \{\g \Omega_{GG},\g \} )
- \Omega_{GG} (\g|\g)^{-1} (\g | \alpha \{\g, \g \}) \right] \nonumber \\
&+ &  \alpha_s \{ \f \Omega_{FG}, \g \} \quad .
\end{eqnarray}
Eqs.(\ref{equ-A-31}) and (\ref{equ-A-32}) together show that $L\gt$ has
the desired structure
\be
\la{equ-A-33}
L\gt = \tilde{V}^{(1)} \gt + \tilde{V}^{(2)} \{\gt, \gt \} \quad ,
\ee
where the matrix elements of $\tilde{V}^{(1)}$ and $\tilde{V}^{(2)}$ can be
taken from (\ref{equ-A-31}) and (\ref{equ-A-32}). A special feature of
$\tilde{V}^{(2)}$ is, that it just couples $\f$ and $\g$, but no
couplings $\f\f$ occur.

Having shown that the condition (\ref{equ-i-1}) holds, 
we use Eq.(\ref{equ-ii-17}) to expand the
residual forces $\tilde{f}$.
As
\be
\la{equ-A-34}
\tilde{V}^{(2)} \sim  \Omega_{FG} \sim V^{(2)} \sim \epsilon
\quad ,
\ee
the interaction $\tilde{V}^{(2)}$ still is of first order in $\epsilon$.
Expanding $\tilde{f}$ we want to keep fixed $\Xi(t)$ and $\Gt$ in the sense
of section A.2. Therefore we must express the correlations $\tilde{\Xi}(t)$ in 
terms of $\Xi(t)$ and $\Gt$. From the definitions of $\tilde{\Xi}$ and
(\ref{equ-A-15}) and (\ref{equ-A-16}) one obtains
\begin{eqnarray}
\la{equ-A-35}
\Xi_{GG}(t) & = & \Xi(t) \nonumber \\
\Xi_{FG}(t) & = & \Gt i \oo \otimes \Xi(t) \nonumber \\
\Xi_{GF}(t) & = & \Xi(t) \otimes i \Omega^{(l)} \Gamma(t) \nonumber \\
\Xi_{FF}(t) & = & \Gt \left( 1 - \otimes \oo 
\Xi(t) \Omega^{(l)} \otimes \Gt \right) \quad .
\end{eqnarray}
This means that fixing $\Xi(t)$ and $\Gt$ is equivalent to fixing the
matrices $\Xi_{GG}, \Xi_{GF}, \Xi_{FG}$, but due to $\oo, \Omega^{(l)}$ in
(\ref{equ-A-35}), the submatrices $\Xi_{FG}$ and $\Xi_{GF}$ are to be
treated as first order terms, whereas $\Xi_{FF}$ has a zeroth and a second
order contribution. 
Hence expanding $\tilde{f}$ in Eq.(\ref{equ-ii-21})
up to first order yields
\be
\la{equ-A-36}
\tilde{f} = \qt \tilde{V}^{(2)} \{ (\gt \tilde{\Xi})^0, 
(\gt \tilde{\Xi} )^0 \} 
+ {\cal O} (\epsilon^2)
\ee
where
\be
\la{equ-A-37}
(\gt \tilde{\Xi})^0_G  =  \g \Xi_{GG}, \qquad
(\gt \tilde{\Xi})^0_F = \f \Gt \quad .
\ee
Explicitly the residual forces (\ref{equ-A-36}) read
\be
\la{equ-A-39}
\tilde{f} = \left( \begin{array}{c}
f_G \\f_F \end{array}  \right)
=
\left( \begin{array}{c}
0  \\
 \qt  \alpha_s 
\{ \f \Gt i \Omega_{FG}, \g \Xi_{GG}(t) \}
\end{array} \right) \quad .
\ee
The selfconsistent equations for $\xt (t)$ are found from $\tilde{\Omega}$
and $(\tilde{f}|\tilde{f}(t))$ to be
\begin{eqnarray}
\la{equ-A-40}
\xd_{GG} & = & \Xi_{GG} i\Omega_{GG} +  \Xi_{GF} i \Omega_{FG} \nonumber \\
\xd_{GF} & = & \Xi_{GG} i \Omega_{GF} + \Xi_{GF} i \Omega_{FF}  - \Xi_{GF} 
\otimes \gamma_{FF} \nonumber \\
\xd_{FG} & = & \Xi_{FG} i \Omega_{GG} +  \Xi_{FF} i \Omega_{FG} \nonumber \\
\xd_{FF} & = & \Xi_{FG} i \Omega_{GF} + \Xi_{FF} i \Omega_{FF} - \Xi_{FF}
\otimes \gamma_{FF}
\end{eqnarray}
where
\be
\la{equ-A-40a}
\gamma_{FF}(t)  =  \ff \left( 
i L \f |\qt  \alpha_s \{ \f \Gt i \Omega_{FG},
\g \Xi_{GG}(t)\}  \right) \otimes \Gt \quad .
\ee
The approximation for the Heisenberg dynamics for $\g (t)$ are obtained from
\be
\la{equ-A-41}
\gt(t) = \gt \xt(t) + \tilde{f}(t) \otimes \xt(t)
\ee
to yield
\be
\la{equ-A-42}
\g (t) = \g \Xi_{GG}(t) + \f \Xi_{FG}(t) 
+ \qt  \alpha_s \{ \f \Gt i \Omega_{FG}, \g \Xi_{GG}(t)\} 
\otimes \Xi_{FG}(t) \quad .
\ee

Now the results of sections A.2 and A.3 can be compared. One sees that the
Heisenberg dynamics (\ref{equ-A-25}) of section A.2 
agree with (\ref{equ-A-42})
of section A.3: Inserting the relations (\ref{equ-A-35}) into (\ref{equ-A-25})
one obtains the results (\ref{equ-A-42}). Furthermore, using the relations
(\ref{equ-A-35}) as a definition for the left hand sides, one can convert 
the equations of motion (\ref{equ-A-23}) and (\ref{equ-A-24}) into the
Eqs.(\ref{equ-A-40}) for $\xt$. Thus we have demonstrated that the approaches
of sections A.2 and A.3 
are equivalent. It is clear, however, that one can also
expand $\tilde{f}$ with $\xt (t)$ being fixed and taking $\Xi_{GF}, \Xi_{FG}$
of first order, which replaces $\Gt$ in (\ref{equ-A-37}) by $\Xi_{FF}$.
Alternatively, one can directly use the expansion of section 2.2 leading to
$\tilde{f} = \qt \vv (\tilde{\g} \xt, \tilde{\g} \xt )$, 
c.f. Eq.(\ref{equ-A-36}).
Such expansions sum up further higher order contributions and would include
the results of section A.2. The spin model of section 3.1 is treated in this
way.

We conclude this section making some remarks upon the static correlations
which occur in the treatment with the extended set $\gt$. Let the approximate
$\tilde{f}$ be a linear combination of 
$\qt (\g \f)$ (e.g. Eq.(\ref{equ-A-39})),
and take the commutator forms for $\tilde{\Omega}$ and 
$(\tilde{f}| \tilde{f}(t))$, then
the correlation functions $\xt (t)$ depend on $(\gt |\gt )$ and 
$\langle \gt \gt \rangle$, and $(\gt |\gt \gt)$, similar to the case of 
Eq.(\ref{equ-ii-35}). 
The essential point however is, that owing to the commutators
$[\gt, \gt ]$, $(\gt | \gt \gt ) $ only enter 
as combinations $(\g |\g \f )$
 and $(\f |\g \f )$. 
Treating the correlations along the lines of section 2., with
exact equations relating $(\gt |\gt ), \langle \gt \gt \rangle$ to
$(\g |\g \g ), (\g |\g \f ), (\f |\g \f )$, and using the approximations for
$(\g |\g \g ), (\g |\g \f ), (\f |\g \f ),$ one sees that 
the system is closed, if
$\langle \f \g \f \rangle$ is included.
\section{$n$--th order approximation and exact solution for the spin model
of section 3.1}
\subsection{Selfconsistent equations for $\Xi^{(n)}$}
Take the orthogonal set
$\gt = \{ \vec{\g}_0, \vec{\g}_{q,1}, \ldots, \vec{\g}_{q,n}; q \neq 0 \}$
defined by (\ref{equ-iii-29}), 
then the time derivatives of $\vec{\g}_{q,\nu}$ read
\bea
\la{equ-B-1}
iL  \vec{\g}_0 & = & 0 \nonumber \\
iL \vec{\g}_{q, 1} &=& \vec{\g}_{q, 2} \nonumber\\
\vdots & & \nonumber\\
iL  \vec{\g}_{q,\nu} & = & - c_\nu  \vec{\g}_{q,\nu-1} 
+ \vec{\g}_{q,\nu+1} \qquad \nu = 2, \ldots,
n-1 \nonumber \\
\vdots & & \nonumber\\
iL  \vec{\g}_{q, n} & = & - c_n \; \; \vec{\g}_{q, n-1} + \frac{J}{\sqrt{N}} 
\left( \vec{\g}_0 \times \vec{\g}_{q, n} - \vec{\g}_{q, n} \times 
\vec{\g}_0 \right)
\eea
with $c_\nu$ defined by Eq.(\ref{equ-iii-37}).
This means that in the interaction
\be
\la{equ-B-3}
L\gt = \tilde{V}^{(1)} \gt + \tilde{V}^{(2)} \{\gt , \gt \}
\ee
just matrix elements $\tilde{V}^{(2)}_{n,0 n}$ and
$\tilde{V}^{(2)}_{n,n 0}$ are different from zero. 
Therefore the approximation (\ref{equ-ii-20}) 
with $\epsilon = J$ for the residual forces yields 
\bea
\la{equ-B-4}
f^{\alpha}_{q, \nu}(t) & = & 0  \qquad \nu = 1, \ldots, n - 1 \nonumber \\
f^{\alpha}_{q, n}(t) & = & \qt \frac{J}{\sqrt{N}} 
(\vec{\g}_0 \times 
\vec{\g}_{q, n} - \vec{\g}_{q, n} \times \vec{\g}_0)^{\alpha}
\Xi_{nn}(t) + {\cal O} (\epsilon^2) \nonumber \\
& = & \g^{\alpha}_{q, n+1} \Xi_{nn}(t) + {\cal O}(\epsilon^2), 
\qquad \alpha = x \mbox{ or } y, z
\eea
leading to the memory matrix 
\be
\la{equ-B-5}
\gamma_{\nu \mu}(t)  =  \frac{(f^{\alpha}_{q, \nu}|f^{\alpha}_{q, \mu}(t)) }
{(\ga |\ga)}
 =  \delta_{\nu n} \delta_{\mu n} c_{n+1} \Xi_{nn}(t) + {\cal O} (\epsilon^3)
\quad .
\ee
The equations of motion of the matrix $\Xi_{\nu \mu}(t)$ (\ref{equ-ii-9})
follow from (\ref{equ-ii-7}) with
\be
\la{equ-B-6}
i\Omega_{\nu \mu} = - c_{\mu} \delta_{\nu, \mu-1} + \delta_{\nu, \mu+1}
\ee
and (\ref{equ-B-5}) for $\gamma_{\nu \mu}(t).$ In the approximation for the
Heisenberg dynamics (\ref{equ-ii-23}) 
the matrix elements $\Xi_{\nu 1}, 
\nu=1, \ldots, n$
and $\Xi_{n n}$ enter. We therefore
regard the Laplace transforms of $\Xi_{ \nu 1}$ and
$\Xi_{n\mu}$ which can be written as
\be
\la{equ-B-7}
\begin{array}{lclclcl}
s \Xi_{11}  & + & c_2 \Xi_{21} & &          & = & 1  \\
s \Xi_{21}  & + & c_3 \Xi_{31}  &-&  \Xi_{11} & = & 0  \\
\vdots      &   &             & &           &   &   \\
s \Xi_{\nu1} & + & c_{\nu+1} \Xi_{\nu+1, 1} & - & \Xi_{\nu-1, 1} & 
= & 0 \qquad \nu = 2, \ldots, n-1
\end{array} 
\ee
and
\be
\la{equ-B-8}\begin{array}{lclclcl}
s \Xi_{n1} & - & \Xi_{n2} &   &              & = & 0  \\
s \Xi_{n2} & - & \Xi_{n3} & + & c_2 \Xi_{n1} & = & 0  \\
\vdots     &   &          &   &              &   &  \\
s \Xi_{n\nu} & - & \Xi_{n\nu+1} & + & c_{\nu} \Xi_{n\nu-1} 
& = & 0 \qquad \nu=2, \ldots, n-1 \\
s \Xi_{nn} & + & c_{n+1} (\Xi_{nn})^2 & + & c_n \Xi_{nn-1} & =&  1 \end{array}
\quad .
\ee
Eqs.(\ref{equ-B-7}) and (\ref{equ-B-8}) provide the selfconsistent solutions.
The coefficients $c_{\nu}$ can be found from the orthogonality of the
$\g_{q, \nu}$ to yield
\bea
\la{equ-B-9}
c_1  & =& 1 \nonumber \\
c_{\nu} & =&  - \frac{\varphi_{\nu} \varphi_{\nu-2}}{\varphi^2_{\nu-1}},
\qquad
\nu > 1 \nonumber \\
\varphi_0 & =& 1, \quad \varphi_{\nu} =
\left|
\begin{array}{cccc}
\alpha_1 & \alpha_2 & \ldots & \alpha_{\nu} \\
\alpha_2 & \alpha_3 & \ldots & \alpha_{\nu+1} \\
\vdots & & & \\
\alpha_{\nu} & \alpha_{\nu+1} & \ldots & \alpha_{2\nu-1} 
\end{array}
\right| \quad
\begin{array}{rcl}
\alpha_{2\nu} & = & 0 \\
\alpha_{2\nu+1} & = & (-1)^{\nu} m_{2\nu} \\
\nu & \geq  & 0 , \\
\end{array}
\eea
where $m_{2\nu}$ denotes the exact $2\nu$-th moment
\be
\la{equ-B-10}
m_{2\nu} = \frac{(\gaq |(L)^{2\nu} \gaq)}{(\gaq | \gaq)} \quad ,
\ee
which can be expressed in the thermodynamic limit $N \rightarrow \infty$ by
the second moment (see (\ref{equ-B-31}))
\be
\la{equ-B-11}
m_{2\nu} = (m_2)^{\nu} \frac{2\nu+1}{6^{\nu}} \frac{(2\nu)!}{\nu!}
\quad .
\ee
\subsection{Explicit Solution of $\Xi^{(n)}(s)$}
The system (\ref{equ-B-7}) can be solved in terms of $\Xi_{11} = \Xi^{(n)}$. 
With help of the polynominals $A_{\nu}(s)$ and $B_{\nu}(s)$ defined by
Eq.(\ref{equ-iii-36}) one directly finds the result (\ref{equ-iii-33}).
For the solution of the system (\ref{equ-B-8}) we first take the equations
for $\nu=1, \ldots, n-1$ expressing $\Xi_{n \nu+1} $ by $\Xi_{n1}$ which gives
\be
\la{equ-B-14}
\Xi_{n \nu+1} = B_{\nu} \Xi_{n1} \quad \nu = 1, \ldots, n-1 \quad .
\ee
With help of (\ref{equ-B-14}) and
(\ref{equ-iii-36}) and
\be
\la{equ-B-15}
\Xi_{nn-1} = \frac{B_{n-2}}{B_{n-1}} \Xi_{nn}
\ee
the last equation of (\ref{equ-B-8}) 
can be converted into a quadratic equation for $\Xi_{nn}$
\be
\la{equ-B-17}
c_{n+1} (\Xi_{nn})^2 + \frac{B_n}{B_{n-1}} \Xi_{nn} - 1 = 0\quad ,
\ee
which has the solution
\be
\la{equ-B-18}
\Xi_{nn} = \frac{ \displaystyle -\frac{B_n}{B_{n-1}} +
\sqrt{\left( \frac{B_n}{B_{n-1}} \right)^2 + 4 c_{n+1}}}{2 c_{n+1}}
\quad .
\ee
The other solution is not a Laplace transform of a function $\Xi_{nn}(t)$
regular at $t=0$.
Combining Eq.(\ref{equ-iii-33}) for $\nu=n$
with (\ref{equ-B-14}) for $\nu=n-1$ 
gives the solution (\ref{equ-iii-35}) of section 3.1 for $\Xi^{(n)}(s)$
which can also be written in terms of $\Xi_{nn}$ as
\be
\la{equ-B-20}
\Xi^{(n)}(s) = \frac{A_n + c_{n+1} \Xi_{nn} A_{n-1}} 
{B_n + c_{n+1} \Xi_{nn} B_{n-1} }
\ee
if, following from Eq.(\ref{equ-iii-36}), the relation
\be
\la{equ-B-21}
A_n B_{n-1} - A_{n-1} B_n = (-1)^{n+1} c_1 \cdots c_n
\ee
is used.

The analytic properties of $\Xi^{(n)}(s)$ follow from
$\Xi_{nn}(s)$ and the representation (\ref{equ-B-20}). First one proves
that $\Xi_{nn}(s)$ is holomorphic for $\mbox{Re} s > 0$, and it holds
\be
\la{equ-B-22}
\mbox{Re} \Xi_{nn}(s) > 0 \quad\mbox{for}\quad \mbox{Re} s > 0 
\quad .
\ee
To see this, we take the definitions of the polynominals $B_{\nu}(s)$ writing
\be
\la{equ-B-23}
\frac{B_{\nu}}{B_{\nu-1}} = s + \frac{c_{\nu}}{\frac{B_{\nu-1}}{B_{\nu-2}}},
\quad c_{\nu} > 0
\ee
which by induction leads to the result
\be
\la{equ-B-24}
\mbox{Re} \frac{B_{\nu}}{B_{\nu-1}} > 0 , \quad \mbox{Re} s > 0
\quad .
\ee
Together with the property, that the $B_{\nu}(s)$ have no zeros for
$\mbox{Re} s > 0$ \cite{9}, 
the mapping (\ref{equ-B-18}) of $B_n/B_{n-1}$
onto $\Xi_{nn}$ completes the proof. For $\Xi^{(n)}(s)$ in Eq.(\ref{equ-B-20})
we use the recurrence relations (\ref{equ-iii-36}) writing
\be
\la{equ-B-24a}
\frac{A_n + c_{n+1} \Xi_{nn} A_{n-1}} 
{B_n + c_{n+1} \Xi_{nn} B_{n-1} } =
\frac{A_{n-1} + \frac{c_{n}}{s + c_{n+1}  \Xi_{nn}} A_{n-2}}  
{B_{n-1} + \frac{c_{n}}{s + c_{n+1}  \Xi_{nn}} B_{n-2}}  
\ee
where from (\ref{equ-B-22}) it follows that
\be
\la{equ-B-24b}
\mbox{Re}\frac{1}{s+c_{n+1} \Xi_{nn}(s)} >0 \quad \mbox{for}
\quad \mbox{Re}s>0 \quad .
\ee
Iterating this procedure one sees that (\ref{equ-B-22}) 
induces $\Xi^{(n)}(s)$
to be holomorphic for $\mbox{Re}s>0$ and to have the property
\be
\la{equ-B-25}
\mbox{Re} \Xi^{(n)} (s) > 0 \quad\mbox{for}\quad \mbox{Re} s > 0 \quad .
\ee
As a last point it is not difficult to see that for $s$
approaching the imaginary
axis, $\mbox{Re} s \rightarrow +0$, 
the function $\Xi(s)$ keeps to be finite.
\subsection{Exact solution $\Xi(t)$}
To derive the exact solution for the dynamic correlation function 
\be
\la{tmpxxx}
\Xi(t) = \frac{\Tr  \gam \gaq (t)}{\Tr  \gam \gaq}
\ee
we start with an
exact differential equation of second order 
for the Heisenberg dynamics of $\g_q(t)$
\be
\la{equ-B-26}
\frac{d^2\gv_q}{d t^2}  + 2 i \frac{J}{\sqrt{N}}
\frac{d\gv_q}{d t} + 4\left(\frac{J}{\sqrt{N}}\right)^2
(\vec{S}_0 \cdot
\vec{S}_0) \gv_q(t) = 0, \quad q \neq 0 \quad .
\ee
In the Hilbertspace of our model ($N$ spins 1/2) we use now a partition of
the identity operator into operators $P_S$ projecting onto the eigenspaces
of $\vec{S}_0 \cdot \vec{S}_0$
leading immediately to
\bea
\la{equ-B-27}
 \frac{d^2}{d t^2} \Tr  \gam P_S \gaq(t) &+&
2 i \frac{J}{\sqrt{N}} \frac{d}{d t} \Tr  \gam P_S \gaq(t) \nonumber\\
&+& 
4 \left( \frac{J}{\sqrt{N}} \right)^2 S(S+1) 
\Tr  \gam P_S \gaq(t)  = 0
\eea
with the solution
\bea
\la{equ-B-28}
& &\Tr  \gam P_S \gaq(t) \nonumber\\
& = & 
\left( \frac{S+1}{2S+1} \Tr  \gam P_S \gaq - \frac{i/2}{2S+1}
\frac{\sqrt{N}}{J}
\Tr  \gam P_S 
\dot{G}_q^{\alpha}(0)
\right) e^{2iSt J/\sqrt{N}} \nonumber \\
&+&  \left( \frac{S}{2S+1} \Tr  \gam P_S \gaq + 
\frac{i/2}{2S+1} \frac{\sqrt{N}}{J}
\Tr  \gam P_S 
\dot{G}_q^{\alpha}(0)
\right) e^{-2i(S+1) t J/\sqrt{N}} 
\quad .
\eea
By calculating the traces which are independent of $q$ and $\alpha = x$
or $y, z$, with
\be
\la{equ-B-29}
\Tr  P_S = \frac{(2S+1)^2}{N/2 + S+1} \left( \begin{array}{c}
 N \\ N/2+S \end{array} \right) \quad ,
\ee
and evaluating the sums we arrive at
\be
\la{equ-B-30}
\Xi_{(N)}(t) = N\left( \cos \frac{J}{\sqrt{N}} t \right)^{N-2} 
\left(\frac{N+1}{N} \cos^2 \left( \frac{J}{\sqrt{N}} t \right)
-1 \right)
\ee
or, taking the thermodynamic limit $N \rightarrow \infty$
\be
\la{equ-B-31}
\Xi(t) = \lim_{N\rightarrow \infty} 
\Xi_{(N)}(t) = e^{- J^2 t^2 /2} (1-J^2 t^2) \quad .
\ee
\section{Dynamic and static correlations of a Heisenberg ferromagnet}
\subsection{Selfconsistent equations in lowest order}
By the approximation discussed in section 3.2 
the dynamic correlation functions 
$\Xi^z(t)$ and $\Xi^{\pm}(t)$
 of a Heisenberg ferromagnet depend on the following static correlations
$(\dsz |\dsz)$, $\langle \dsz \dsz \rangle$, $(\shpm | \shpm)$,
$\langle \shm \hat{S}^z \rangle$, $( \dsz | \shm  \shp)$, 
$(\shpm | \dsz \shpm)$, 
$\langle \shm \shp|\dsz \rangle$, $(\shpm |
\shm \shp \shpm)$ (see Eqs. (\ref{equ-iii-68})--(\ref{equ-iii-77})).

In this appendix we want to point out now, that one can close the dynamic
equations (\ref{equ-iii-68})--(\ref{equ-iii-74}) of section 3.2 by a set of
relations determining the static correlations in terms of the dynamic ones. The
procedure is analogous to the the general scheme discussed in section 2.5,
the only difference is, that the iteration with 
$V_0^{(2)} + \epsilon V_1^{(2)}$ 
has led to residual forces, which have contributions of the form $\g\g\g$.
Therefore there appear higher--product static correlations, e.g.
$(\shpm | \shm \shp \shpm)$, and the set of static quantities has to be 
extended in order to find a closed system of equations.

Again we start with exact relations between $(\g|\g)$ and $(\g|\g\g)$ on the
one hand, and between $(\g|\g)$ and $\langle \g^{\dagger} \g \rangle$
on the other hand (compare Eqs.(\ref{equ-ii-36}) and 
(\ref{equ-ii-37}) of
section 2.5)\footnote{For the longitudinal case the relation analogous to Eq.
(\ref{equ-C-3}) would read 
\be
\la{equ-C-1a}
 0 = \sum_2 (J_{1-2}-J_2) (\dsz_1|\shm_2 \shp_{1-2})
\ee
and would not supply any further information, 
because by symmetry Eq.(\ref{equ-C-1a})
is identically fulfilled. We therefore have replaced 
Eq.(\ref{equ-C-1a}) by
(\ref{equ-C-1}), following from
\begin{displaymath} 
\sum_2 (\dsz_1|\vec{S}_2 \cdot \vec{S}_{1-2}) = 0
\end{displaymath}}
\bea
(\dsz_1|\dsz_1) &=& 
-  \frac{2}{(2+\epsilon)N}  \sum_2 
(\dsz_1|\shm_2 \shp_{1-2})
-  \frac{\epsilon}{(2+\epsilon)N} \sum_{2} (\dsz_1|\dsz_2 \dsz_{1-2})
\la{equ-C-1} \\
N&=&2 \bt (J_0-J_1)(\shpm_1| \shpm_1) - \epsilon \frac{2\bt}{N} \sum_{2}
(J_{1-2}-J_2) (\shpm_1 | \dsz_2 \shpm_{1-2})
\la{equ-C-3} \\
\langle \dsz_{-1} \dsz_1\rangle &=& (\dsz_1 | \dsz_1) \int d\omega 
\frac{\bt \omega}{e^{\bt \omega}-1} \Xi^z_1 (\omega)
\la{equ-C-2}\\
\langle \shm_{-1} \shp_1 \rangle &=& (\shp_1 | \shp_1) \int d\omega 
\frac{\bt \omega}{e^{\bt \omega}-1} \Xi^+_1 (\omega) \quad .
\la{equ-C-4} 
\eea

Using now the approximation (\ref{equ-iii-61}) and (\ref{equ-iii-62})
in Eqs.(\ref{equ-iii-64}) and (\ref{equ-iii-65}) of section 3.2 for the
Heisenberg dynamics of $\dsz(\omega)$ and $\shp (\omega)$
\bea
\dsz_1 (\omega) &=&  \dsz_1 \Xi^z_1 (\omega) - \frac{2i}{N} \sum_{2}
Q  \shm_2 \shp_{1-2} (J_{1-2} - J_2) (
\phi_{12} \otimes \Xi^z_1) (\omega)
\la{equ-C-5} \\
\shp_1(\omega) &=& \shp_1 \Xi^+_1(\omega)  +  \epsilon \frac{2i}{N}
\sum_{2} (J_{1-2} - J_2) Q \left\{ 
\vphantom{\frac{2i}{N}}
\dsz_2 \shp_{1-2}
\left( (\Xi^z_2 \Xi^+_{1-2}) \otimes \Xi^+_1\right) (\omega) -
\right . \nonumber \\
& - & \left. \frac{2i}{N} \sum_{3} Q (\shm_3 \shp_{2-3}) 
\shp_{1-2} (J_{2-3} - J_3) 
\left[ \left( \left(\phi_{23} \otimes \Xi^z_2\right) 
\Xi^+_{1-2} \right)
\otimes \Xi^+_1 \right]
(\omega) \right\} \quad ,
\la{equ-6}
\eea
the equations analogous to Eqs.(\ref{equ-ii-28}) and (\ref{equ-ii-29}) of
section 2.5 explicitly read
\bea
(A|\dsz_1) &=& a_1 \langle [ S^z_1, A^{\dagger}]\rangle + \sum_{2} b_{12}
\langle [\shm_2 \shp_{1-2}, A^{\dagger}]\rangle
\la{equ-C-7}\\
(A|\shp_1) &=& c_1 \langle [\shp_1, A^{\dagger}]\rangle \nonumber\\
&+&  \epsilon
\sum_{2} d_{12} \langle [\dsz_2 \shp_{1-2}, A^{\dagger}]\rangle 
+  \epsilon \sum_{2,3} e_{123} \langle[\shm_3 \shp_{2-3} \shp_{1-2},
A]\rangle
\la{equ-C-8}\\
\langle A \dsz_1\rangle &=& \hat{a}_1 \langle [S^z_1, A]\rangle + \sum_{2}
\hat{b}_{12} \langle [\shm_2 \shp_{1-2}, A]\rangle
\la{equ-C-9}\\
\langle A \shp_1 \rangle &=&  \hat{c}_1 \langle [ S^z_1, A]\rangle  
\nonumber\\
& +& 
\epsilon \sum_{2} \hat{d}_{12} \langle [\dsz_2 \shp_{1-2}, A]\rangle
+ \epsilon \sum_{2,3} \hat{e}_{123} \langle 
[ \shm_3 \shp_{2-3} \shp_{1-2}, A]\rangle
\la{equ-C-10}
\eea
with coefficients
\bea
a_1 &=& \int \frac{d\omega}{\bt \omega} \Xi^z_1 (\omega)
- \sum_{2} \frac{(\dsz_1|\shm_2 \shp_{1-2})}{(\dsz_1| \dsz_1)} b_{12}
\la{equ-C-11}\\
b_{12} &=& - \frac{2i}{N} (J_{1-2} - J_2) \int 
\frac{d\omega}{\bt \omega} (\phi_{12} \otimes \Xi^z_1) (\omega)
\la{equ-C-12}\\
c_1 &=& \int \frac{d\omega}{\bt \omega} \Xi^+_1(\omega) -  \epsilon \sum_{2}
\frac{(\shp_1|\dsz_2 \shp_{1-2})}{(\shp_1|\shp_1)} d_{12}  
+  \epsilon \sum_{2,3} \frac{(\shp_1 | \shm_3 \shp_{2-3}
\shp_{1-2})}{(\shp_1|\shp_1)}e_{123}
\la{equ-C-13}\\
d_{12} &=&   \frac{2i}{N} (J_{1-2}-J_2)\int \frac{d\omega}{\bt\omega} \left\{
(\Xi^z_2 \otimes \Xi^+_{1-2}) \otimes \Xi^+_1 \right\} (\omega)
-  \sum_{3} \frac{(\dsz_2|\shm_3 \shp_{2-3})}{(\dsz_2|\dsz_2)} e_{123}
\la{equ-C-14}\\
e_{123} &=& -\left(\frac{2i}{N}\right)^2 (J_{1-2}-J_2) (J_{2-3} -J_3)
\int \frac{d\omega}{\bt\omega} 
\left\{ \left( (\phi_{23} \otimes \Xi^z_2)\Xi^+_{1-2} \right) 
\otimes \Xi^+_1 \right\}
(\omega) \quad,
\la{equ-C-15}
\eea
and analogous expressions for the coefficients $\hat{a}_1, \hat{b}_{12},
\hat{c}_1, \hat{d}_{12}$ and $\hat{e}_{123}$. The only difference is, that
the denominators $\bt\omega$ in Eqs.(\ref{equ-C-11})--(\ref{equ-C-15})
are replaced by $(e^{\bt\omega}-1)$.

We now use\footnote{For the transverse case 
it suffices to regard correlations like $(\shm_1|\ldots \shm_1)$. 
The correlations $(\shp_1| \ldots \shp_1)$ follow from
symmetry arguments, e.g. $(\shp_1|\shp_1) = (\shm_1|\shm_1)$.}
Eq. (\ref{equ-C-7}) for
\be
A  = 
\shm_{1-2} \shp_2
\la{equ-C-16} \quad ,
\ee
Eq. (\ref{equ-C-8}) for
\be
\la{equ-C-18}
A = \left\{ \begin{array}{l}  
\shp_{1-2} \dsz_2 \\
\shp_{1-2} \shm_{2-3} \shp_3
\end{array}
\right. \quad ,
\ee
Eq. (\ref{equ-C-9}) for
\be
A = \left\{ \begin{array}{l}
\shm_{-2} \shp_{2-1}\\ 
\dsz_{-2} \dsz_{2-1} \\
\shm_{-3} \shp_{3-2} \dsz_{2-1} \\
\shm_{-4} \shp_{4-3} \shm_{3-2} \shp_{2-1} 
\end{array} \right. \quad ,
\la{equ-C-19}
\ee
and last not least
Eq. (\ref{equ-C-10}) for  
\be
\la{equ-C-23}
A = \shm_{-3} \shp_{3-2} \shm_{2-1} \quad ,
\ee
and evaluate the commutators on the right hand side with help of
\bea
\left[ \shp_1, \shm_2 \right] 
&=& N \delta_{1,-2} + \epsilon \dsz_{1+2} \la{equ-C-24}\\
\left[ \dsz_1, \shpm_2\right] &=& \pm \shpm_{1+2} \la{equ-C-25}
\quad .
\eea
Then it is easy to see, that Eqs.(\ref{equ-C-1})--(\ref{equ-C-4}) together
with Eqs.(\ref{equ-C-16})--(\ref{equ-C-23}) constitute a system of 12
equations determing the 12 static correlations $(\dsz|\dsz)$, 
$\langle\dsz \dsz\rangle$, $(\shm|\shm)$, $\langle \shm \shp \rangle$, 
$(\dsz|\shm \shp)$, $(\shm|\dsz \shm)$,
$\langle \shm \shp \dsz \rangle$, $(\shm| \shm \shp \shm)$,
$\langle \dsz \dsz \dsz \rangle$, $(\shm \shp \shm \shp \rangle$,
$\langle \shp \shp \dsz \dsz \rangle$, 
$\langle  \shm \shp \shm \shp \dsz \rangle$ 
in  terms of the dynamic correlation functions $\Xi^z(t)$ and $\Xi^{\pm}(t)$.
Thus together with the equations of motion (see Eqs.(\ref{equ-iii-68}) and
(\ref{equ-iii-69}) of section 3.2)
\bea
\frac{d}{d\tau} \Xi^z_1 &=& - \gamma^{\parallel}_1 \otimes \Xi^z_1
\la{equ-C-26}\\
\frac{d}{d\tau} \Xi^{\pm}_1 &=& \mp i \omega_1 \Xi^{\pm}_1 - \gamma^{\perp}_1
\otimes \Xi^{\pm}_1
\la{equ-C-27}
\eea
with $ \gamma^{\parallel}_1$ and $\gamma^{\perp}_1$ 
given by Eqs.(\ref{equ-iii-73}) and (\ref{equ-iii-74}) of section 3.2, 
we have found the desired closed system of 
equations coupling static and dynamic correlations.
\subsection{Zeroth order solution}
The easiest way to find the zeroth order contributions in
$\epsilon$ to the dynamic-- and static correlations of a Heisenberg ferromagnet
is to use perturbation theory to the Heisenberg equation of motion 
(\ref{equ-iii-57}) and (\ref{equ-iii-58}) of section 3.2 for $\dsz(\tau)$ and 
$\shpm(\tau)$. For demonstration purposes we want to sketch, how
these simplest approximations to the correlations can be found from the
zeroth order solution of the rather complicated coupled system of equations
constructed in appendix C.1

First the zeroth order of the frequency $\omega_1^{(0)}$ 
(see Eq.(\ref{equ-iii-72})
of section 3.2) can be found from Eq.(\ref{equ-C-3}) to yield
\be
\la{equ-C-28}
\omega_1^{(0)} =\frac{N}{\bt (\shp_1|\shp_1)^{(0)}}=
2(J_0-J_1)=:\tilde{\omega}_1 \quad .
\ee
Since the correlation function $\gamma^{\bot}(\tau)$ of the transverse
residual force is of second order $\gamma^{\bot}(\tau) \sim \epsilon^2$, 
Eq.(\ref{equ-iii-74}) can easily be integrated
\be
\la{equ-C-29}
\Xi_1^{\pm (0)}(\tau) = e^{\mp i \tilde{\omega}_1 \tau}
\ee
by Eq.(\ref{equ-C-4}) immediately leading to
\be
\la{equ-C-30}
\frac{1}{N} \langle \shm_{-1} \shp_1\rangle^{(0)} 
= \left(e^{\bt\tilde{\omega}_1}-1\right)^{-1}
=: n_1 \quad .
\ee

To find the zeroth order of the longitudinal correlations, we first notice, 
that (\ref{equ-C-1}) leads to
\be
\la{equ-C-31}
\left( \dsz_1|\dsz_1\right)^{(0)}= -\frac{1}{N} \sum_2
\left(\dsz_1|\shm_2 \shpm_{1-2}\right)^{(0)}
\ee
and (\ref{equ-C-16}) to
\be
\la{equ-C-32}
\left(\dsz_1|S_2^- S_{1-2}^+ \right)^{(0)} = N
\left(a_1^{(0)}-b_{1 1-2}^{(0)}\right) \left( n_2 -n_{1-2}
\right)
\ee
with
\bea
a_1^{(0)} &=& \intP \frac{d\omega}{\bt \omega} \Xi_1^{z (0)} (\omega) -
\sum_2 
\frac{(\dsz_1| \shm_2 \shp_{1-2})^{(0)}}
{(\dsz_1| \dsz_1)^{(0)}} 
b_{12}^{(0)} \la{equ-C-33}\\
b_{12}^{(0)} &=& -\frac{i}{N} \left(\tilde{\omega}_2-\tilde{\omega}_{1-2}
\right) \intP \frac{d\omega}{\bt \omega} \left( \phi_{12} \otimes
\Xi_1^z\right)^{(0)} (\omega) \quad . \la{equ-C-34}
\eea
To evaluate the principal value integrals they are expressed by
\be
\la{equ-C-36}
\intP \frac{d\omega}{\omega}  \Psi(\omega) =
\frac{i}{2} \left(\Psi_>(\eta) - \Psi_<(-\eta)\right),\quad
\eta\rightarrow +0
\ee
with the abbreviation
\be
\la{equ-C-35}
\Psi_{\gtrless} (s) := \int_0^{\infty} d \tau e^{\mp s \tau} \Psi(\pm \tau)
\quad .
\ee
The expressions $\Xi_{\gtrless}^{z (0)}(s)$ and $(\phi_{12} \otimes
\Xi_1^z)^{(0)}_{\gtrless}(s)$ can be found from the defining
Eq.(\ref{equ-iii-68}) together with 
(\ref{equ-iii-73}) of
section 3.2 using the zeroth order result (\ref{equ-C-29}) of
$\Xi_1^{\pm}(\tau)$. They depend on the static correlations 
$(\dsz_1|\shm_2 \shp_{1-2})^{(0)}$ and $(\dsz_1|\dsz_1)^{(0)}$.
Thus Eqs.(\ref{equ-C-31}) and (\ref{equ-C-32}) together with the
definitions (\ref{equ-C-33}) and (\ref{equ-C-34}) lead to a coupled system
of equations, which can be solved by using Eq.(\ref{equ-C-36}) to yield
\bea
(\dsz_1|\shm_2 \shp_{1-2})^{(0)} &=& N \frac{n_2-n_{1-2}}{\bt
(\tilde{\omega}_2-\tilde{\omega}_{1-2})}\la{equ-C-37}\\
(\dsz_1|\dsz_1)^{(0)} &=& -\sum_2 \frac{n_2-n_{1-2}}{\bt
(\tilde{\omega}_2-\tilde{\omega}_{1-2})}\la{equ-C-38} \quad .
\eea
This implies
\be
\Xi_1^{z (0)}(\tau) = \left( \sum_2 \frac{n_2-n_{1-2}}{\bt
(\tilde{\omega}_2-\tilde{\omega}_{1-2})}\right)^{-1}
\sum_2 \frac{n_2-n_{1-2}}{\bt
(\tilde{\omega}_2-\tilde{\omega}_{1-2})} e^{i (\tilde{\omega}_2
-\tilde{\omega}_{1-2})\tau} \quad . \la{equ-C-39}
\ee
The results for the expectation values
\bea
\langle \dsz_{-1} \dsz_1\rangle^{(0)} &=& \sum_2 n_2 (1+n_{1-2})
\la{equ-C-40}\\
\langle \shm_{-2} \shp_{2-1} \dsz_1\rangle^{(0)} &=& -N n_2
(1+n_{1-2}) \la{equ-C-41}
\eea
follow from Eqs.(\ref{equ-C-2}) and (\ref{equ-C-19}) with 
Eq.(\ref{equ-C-39}).
\end{appendix}
\end{document}